\documentclass[twocolumn]{aastex63}
\usepackage{amsmath}
\usepackage{graphicx}
\usepackage{amsfonts}
\usepackage{natbib}
\usepackage{color}
\usepackage{epstopdf}
\usepackage{graphicx}
\usepackage{multirow}
\usepackage{longtable}
\usepackage{rotating}
\usepackage{lineno}
\usepackage{appendix}
\usepackage{attachfile}
\usepackage{url}
\usepackage{comment}

\shorttitle{Rapid Variability and Broadband Spectral Modeling in the Flaring Activity of BL Lacertae}
\shortauthors{Xin Chang et al.}

\begin{document}

\title{Rapid Variability and Broadband Spectral Modeling in the Flaring Activity of BL Lacertae}
\correspondingauthor{Dingrong Xiong, Chenxu Liu, Rui Xue}
\email{xiongdingrong@ynao.ac.cn, cxliu@ynu.edu.cn, ruixue@zjnu.edu.cn}

\author[0009-0008-1701-2792]{Xin Chang}
\affiliation{South-Western Institute for Astronomy Research, Yunnan Key Laboratory of Survey Science, Yunnan University, Kunming, Yunnan 650504, People's Republic of China}

\author[0000-0002-6809-9575]{Dingrong Xiong}
\affiliation{Yunnan Observatories, Chinese Academy of Sciences, 396 Yangfangwang, Guandu District, Kunming, 650216, People's Republic of China}

\author[0000-0001-5561-2010]{Chenxu Liu}
\affiliation{South-Western Institute for Astronomy Research, Yunnan Key Laboratory of Survey Science, Yunnan University, Kunming, Yunnan 650504, People's Republic of China}

\author[0000-0003-1721-151X]{Rui Xue}
\affiliation{Department of Physics, Zhejiang Normal University, Jinhua 321004, People's Republic of China}

\author[0000-0001-8920-0073]{Tingfeng Yi}
\affiliation{Key Laboratory of Colleges and Universities in Yunnan Province for High-energy Astrophysics, Department of Physics, Yunnan Normal University, Kunming 650500, People's Republic of China}

\author[0000-0003-2017-9159]{Jia Zhang}
\affiliation{College of Physics and Optoelectronic Engineering, Leshan Normal University, Leshan 614000, People's Republic of China}

\author[0009-0002-7625-2653]{Yu Pan}
\affiliation{South-Western Institute for Astronomy Research, Yunnan Key Laboratory of Survey Science, Yunnan University, Kunming, Yunnan 650504, People's Republic of China}

\author[0009-0006-5847-9271]{Xingzhu Zou}
\affiliation{South-Western Institute for Astronomy Research, Yunnan Key Laboratory of Survey Science, Yunnan University, Kunming, Yunnan 650504, People's Republic of China}

\author[0009-0000-4068-1320]{Xinlei Chen}
\affiliation{South-Western Institute for Astronomy Research, Yunnan Key Laboratory of Survey Science, Yunnan University, Kunming, Yunnan 650504, People's Republic of China}

\author[0000-0001-8278-2955]{YeHao Cheng}
\affiliation{South-Western Institute for Astronomy Research, Yunnan Key Laboratory of Survey Science, Yunnan University, Kunming, Yunnan 650504, People's Republic of China}

\author[0000-0001-6374-8313]{Yuanpei Yang}
\affiliation{South-Western Institute for Astronomy Research, Yunnan Key Laboratory of Survey Science, Yunnan University, Kunming, Yunnan 650504, People's Republic of China}

\author[0000-0002-2510-6931]{Jinghua Zhang}
\affiliation{South-Western Institute for Astronomy Research, Yunnan Key Laboratory of Survey Science, Yunnan University, Kunming, Yunnan 650504, People's Republic of China}

\author[0000-0003-0394-1298]{Xiangkun Liu}
\affiliation{South-Western Institute for Astronomy Research, Yunnan Key Laboratory of Survey Science, Yunnan University, Kunming, Yunnan 650504, People's Republic of China}

\author[0009-0006-1010-1325]{Yuan Fang}
\affiliation{South-Western Institute for Astronomy Research, Yunnan Key Laboratory of Survey Science, Yunnan University, Kunming, Yunnan 650504, People's Republic of China}

\author[0000-0002-8109-7152]{Guowang Du}
\affiliation{South-Western Institute for Astronomy Research, Yunnan Key Laboratory of Survey Science, Yunnan University, Kunming, Yunnan 650504, People's Republic of China}

\author[0009-0005-8762-0871]{Tao Wang}
\affiliation{South-Western Institute for Astronomy Research, Yunnan Key Laboratory of Survey Science, Yunnan University, Kunming, Yunnan 650504, People's Republic of China}

\author[0009-0003-6936-7548]{Xufeng Zhu}
\affiliation{South-Western Institute for Astronomy Research, Yunnan Key Laboratory of Survey Science, Yunnan University, Kunming, Yunnan 650504, People's Republic of China}

\author[0000-0003-1984-3852]{Zhongxiang Wang}
\affiliation{Department of Astronomy, School of Physics and Astronomy, Yunnan University, Kunming 650091, People's Republic of China}

\author[0000-0003-0038-5548]{Sarira Sahu}
\affiliation{Instituto de Ciencias Nucleares, Universidad Nacional Aut\'{o}noma de M\'{e}xico, Circuito Exterior S/N, C.U., A. Postal 70-543, CDMX 04510, M\'{e}xico}

\author[0000-0003-1295-2909]{Xiaowei Liu}
\affiliation{South-Western Institute for Astronomy Research, Yunnan Key Laboratory of Survey Science, Yunnan University, Kunming, Yunnan 650504, People's Republic of China}

\begin{abstract}
We report a multi-wavelength study of two flaring episodes of the blazar BL Lacertae during MJD 60500-60800 (9 July 2024 - 5 May 2025). The source reached a daily-averaged $\gamma$-ray flux of $(1.03 \pm 0.05) \times 10^{-5} \, \mathrm{ph \, cm^{-2} \, s^{-1}}$ ($E > 100$ MeV) on MJD 60588 (5 October 2024). Using orbit-binned data from the Large Area Telescope (LAT) onboard the \textit{Fermi Gamma-ray Space Telescope}, we identify a minimum flux halving timescale of $\tau = 1.33 \pm 0.29$ hr. This constrains the upper limit on the $\gamma$-ray emitting region size to $R \le 2.0 \times 10^{15}$ cm, as well as its distance from the central supermassive black hole to $R_\mathrm{H} \le 5.9 \times 10^{16}$ cm, assuming a Doppler factor of $\delta = 14.8$ derived from the spectral energy distribution (SED) modeling. We find tentative evidence for sub-minute $\gamma$-ray variability with a minimum doubling time of $0.7 \pm 0.2$ min ($p$-value = 0.03). This may originate from an extremely compact region with a size of $R \le 1.8 \times 10^{13}$ cm, suggesting that the emission arises from magnetohydrodynamic substructures, such as plasmoids within a magnetic reconnection zone. Spectral analysis reveals a significant ``softer-when-brighter'' trend ($r = 0.96, p = 4.5 \times 10^{-4}$) during the minute-scale flare peaks, indicating a complex interplay between particle acceleration and radiative cooling. The SED is reproduced using a one-zone leptonic model, in which synchrotron self-Compton (SSC) and external Compton (EC) scattering effectively account for the high-energy emissions. The reduced magnetic field strengths and hard electron injection spectral indices observed during the flaring states suggest enhanced particle acceleration efficiency, possibly associated with relativistic magnetic reconnection.
\end{abstract}


\section{Introduction}\label{sec:intro}
Blazars are traditionally categorized into BL Lac objects and flat-spectrum radio quasars (FSRQs) based on the presence or absence of broad emission lines in their optical spectra. Alternatively, blazars can be classified according to their synchrotron peak frequency ($\nu_{\text{peak, syn}}$): FSRQs typically exhibit low peak frequencies in the infrared, whereas BL Lacs exhibit peaks ranging from radio to X-ray frequencies. Based on $\nu_{\text{peak, syn}}$, blazars are further divided into low-, intermediate-, and high-synchrotron-peaked blazars (LSP, ISP, and HSP), with peak frequencies below $10^{14}$ Hz, between $10^{14}$ and $10^{15}$ Hz, and above $10^{15}$ Hz, respectively \citep{2010ApJ...716...30A}. These two subclasses of blazars typically correspond to different accretion regimes: FSRQs generally feature radiatively efficient, optically thick accretion disks, whereas BL Lac objects are characterized by radiatively inefficient accretion flows. In the latter, the weaker ionizing radiation leads to weak (or absent) broad emission lines \citep{2009MNRAS.396L.105G}. Blazars exhibit pronounced flux and spectral variability across the entire electromagnetic spectrum, manifesting over a wide range of timescales from minutes to years \citep[e.g.,][]{1995ARA&A..33..163W,2008Natur.452..966M,2013MNRAS.436.1530R}. BL Lacertae ($z = 0.069$) \citep{1977ApJ...212L..47M} is the prototype of the BL Lac class. Depending on its synchrotron peak frequency, the source has been classified as either an LSP BL Lac (LBL) \citep{2018A&A...620A.185N} or an ISP BL Lac (IBL) \citep{2020ApJ...892..105A}, and it is also recognized as a TeV-emitting blazar \citep{2019A&A...623A.175M}. Notably, the optical spectrum of BL Lacertae reveals broad $\mathrm{H}\alpha$ and $\mathrm{H}\beta$ emission lines \citep{1995ApJ...452L...5V} with time-variable line fluxes \citep{2010A&A...516A..59C}. The presence of these weak but distinct broad lines suggests that BL Lacertae may share more physical similarities with FSRQs than with lower-luminosity BL Lac objects.

BL Lacertae is renowned for its intense short-term and intraday variability \citep{1989Natur.337..627M}. Between August and November 2019, BL Lacertae exhibited a minimum variability timescale of approximately 30 min at optical wavelengths, whereas the corresponding X-ray variability timescale was significantly longer at 14.5 hr \citep{2020ApJ...900..137W}. More rapid variations have been observed during subsequent flaring states. During the brightest $\gamma$-ray flare on 27 April 2021 (MJD 59331), \citet{2023MNRAS.521L..53A} detected sub-hour variability of $46 \pm 24$ min using orbit binning. Furthermore, during the X-ray flux peak on 6 October 2020 (MJD 59128), a minimum variability timescale of $7.7 \pm 1.6$ min was detected ($4.8\sigma$), while 30 s binned data suggested a potentially shorter timescale of $2.4 \pm 0.9$ min ($2.6\sigma$) \citep{2023MNRAS.521L..53A}. Most notably, \citet{2022MNRAS.509...52D} identified an even shorter variability timescale of 50 s during the same activity peak, thereby constraining the intrinsic size of the emitting region to $R < 1.1 \times 10^{14}$ cm. Minute-timescale variations have also been observed in the $\gamma$-ray band from BL Lacertae. Specifically, \citet{2013ApJ...762...92A} observed a rapid very-high-energy (VHE) flare with an exponential decay time of $13 \pm 4$ min; similarly, \citet{2019A&A...623A.175M} reported a VHE flare with a halving time of $26 \pm 8$ min. On 27 April 2021 (MJD 59331), \citet{2022A&A...668A.152P} reported pronounced $\gamma$-ray variability in light curves with minute-scale temporal binning, identifying flux halving timescales as short as approximately 1 minute. Accompanying this rapid variability, a peak $\gamma$-ray flux (0.1--500 GeV) of $2.0 \times 10^{-5}~\mathrm{ph~cm^{-2}~s^{-1}}$ was recorded, indicating that the $\gamma$-ray emission is highly beamed and likely originates from an exceptionally compact region within the jet. Relativistic particle acceleration may be driven by magnetic reconnection \citep[e.g.,][]{2020MNRAS.494.1817D}, a process in which magnetic energy is dissipated and converted into particle kinetic energy, leading to the observed rapid variability.

The spectral energy distribution (SED) of blazars is characterized by a broadband non-thermal continuum extending from radio to $\gamma$-ray frequencies, exhibiting two distinct emission components. The low-energy component, which peaks between the infrared and X-ray bands, originates from the synchrotron radiation of relativistic electron-positron ($e^{\pm}$) pairs. The high-energy component, peaking in the $\gamma$-ray band, can be attributed to inverse Compton (IC) scattering between $e^{\pm}$ pairs and various soft photon fields within leptonic models \citep{1998MNRAS.299..433F,2010MNRAS.401.1570T}. \citet{2011ApJ...733L..26A} conducted the first multi-wavelength study of BL Lacertae, demonstrating the presence of a broad-line region (BLR) through broadband SED modeling. In addition, \citet{2024MNRAS.528.7587H} showed that external photon fields play a crucial role in the high-energy emission of LBLs and suggested that the $\gamma$-ray emitting regions are likely located outside the BLR but within the dusty torus (DT). Investigations into SEDs of blazars during flaring epochs provide crucial constraints on the underlying particle acceleration mechanisms within relativistic jets.

On 5 October 2024 (MJD 60588), BL Lacertae underwent a major outburst, reaching a peak $\gamma$-ray flux (100 MeV--300 GeV) of approximately $1.03 \times 10^{-5}~\mathrm{ph~cm^{-2}~s^{-1}}$, followed by a subsequent flare on 12 February 2025 (MJD 60718). The physical mechanisms underlying these extreme flares remain subject to significant uncertainty. Exploiting these exceptionally high-significance flaring events, we probe the $\gamma$-ray variability timescales using high-time-resolution light curves and conduct an in-depth analysis of these episodes through broadband SED modeling within a leptonic framework, aiming to elucidate their origin and the dominant radiation processes involved.

The multi-wavelength dataset utilized in this study integrates observations from \textit{Fermi} Large Area Telescope (LAT), \textit{Swift} X-ray Telescope (XRT), \textit{Swift} Ultraviolet Optical Telescope (UVOT), the Zwicky Transient Facility (ZTF), and the 1.6-m Multi-channel Photometric Survey Telescope (Mephisto). The paper is structured as follows: Section \ref{sec_data} details the data acquisition and analytical methodologies; Section \ref{sec_analysis} presents the results of the multi-wavelength temporal and spectral analyses; and Sections \ref{sec_discuss} and \ref{sec_sum} provide the discussion and summary of our findings.

\begin{figure*}
\centering
\includegraphics[scale=.82]{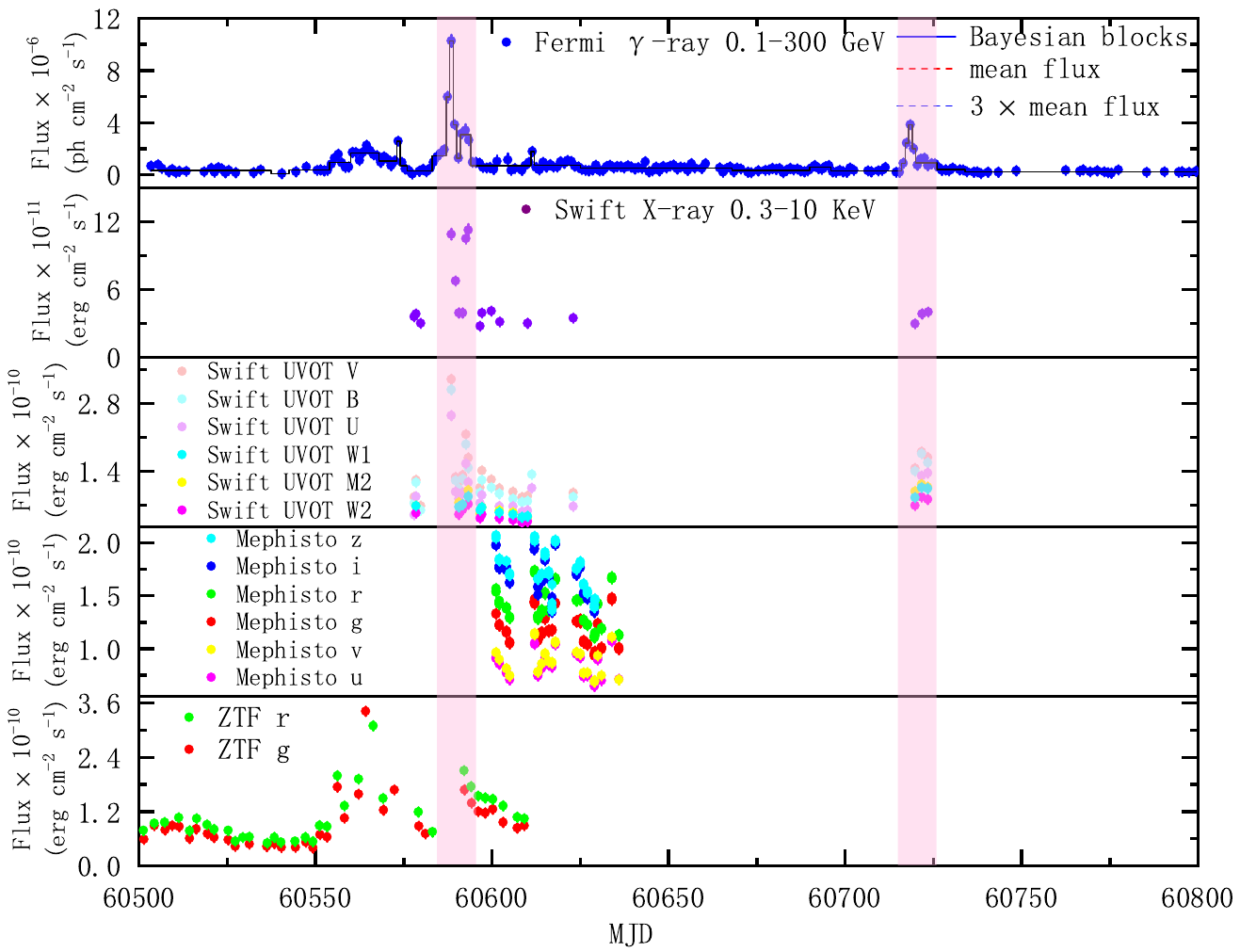}
\caption{Multi-wavelength light curves of BL Lacertae from MJD 60500 to 60800 (9 July 2024 - 5 May 2025). From top to bottom, the panels display: \textit{Fermi}-LAT $\gamma$-ray flux ($0.1\text{--}300$ GeV); \textit{Swift}-XRT X-ray flux ($0.3\text{--}10$ keV); \textit{Swift}-UVOT flux (across $V, B, U, UVW1, UVM2, UVW2$ filters); Mephisto flux in the $u, v, g, r, i$, and $z$ bands; and ZTF flux ($g$ and $r$ bands). The X-ray, UV, and optical data are corrected for Galactic extinction. The data behind this figure are available in machine-readable format in the online journal.}
\label{f_1}
\end{figure*}

\section{Observations and Data Reduction}
\label{sec_data}
\subsection{\textit{Fermi}-LAT}
The \textit{Fermi}-LAT onboard the \textit{Fermi Gamma-ray Space Telescope} was launched by NASA in June 2008 \citep{2009ApJ...697.1071A}. The \textit{Fermi}-LAT is designed to measure the arrival times, directions, and energies of incident $\gamma$-rays, detecting photon events within an energy range from 20 MeV to over 300 GeV \citep{2009ApJ...697.1071A}. On 5 October 2024 (MJD 60588), a significant outburst from the blazar BL Lacertae was detected by the LAT. To investigate its $\gamma$-ray behavior, we analyzed data collected from 9 July 2024 to 5 May 2025 (MJD 60500--60800). This observation period encompasses multiple flares, covering both the low- and high-flux states of the source.

In this study, we utilized LAT Pass 8 data (specifically \texttt{evclass = 128} and \texttt{evtype = 3}) in the energy range from 100 MeV to 300 GeV. A Region of Interest (ROI) with a $15^\circ$ radius was centered on the source position. Analysis was performed following standard procedures using the \textit{Fermi} Science Tools (version 2.2.0). To mitigate contamination from $\gamma$-rays originating from the Earth's limb, a maximum zenith angle cut of $90^{\circ}$ was applied. Good Time Intervals (GTIs) were selected using the recommended screening expression: \texttt{(DATA\_QUAL > 0) \&\& (LAT\_CONFIG == 1)}. The source model was constructed using an XML file including the \texttt{gll\_iem\_v07} Galactic diffuse emission model and the \texttt{iso\_P8R3\_SOURCE\_V3\_v1} isotropic background model. Sources with a test statistic $(\mathrm{TS})<9$ (corresponding to a significance of approximately $3\sigma$; \citet{1996ApJ...461..396M}) were excluded from the analysis. We generated a light curve with one-day binning to characterize the source's variability, as illustrated in Figure \ref{f_1}. For the spectral analysis, a LogParabola model was adopted as the default for BL Lacertae \citep{2023arXiv230712546B}. Model parameters were optimized via an unbinned likelihood analysis using the Python-based tools developed by the \textit{Fermi} collaboration.

\subsection{\textit{Swift}-XRT}
This study utilizes data from the \textit{Swift}-XRT ($0.3\text{--}10.0$ keV) onboard the \textit{Neil Gehrels Swift Observatory}. The dataset spans from MJD 60500 to 60800 (9 July 2024 - 5 May 2025), encompassing 20 \textit{Swift} observations of BL Lacertae. Each unique observation ID corresponds to a single data point in the \textit{Swift}-XRT light curve, as shown in Figure \ref{f_1}. The XRT is a grazing-incidence focusing spectrometer \citep{2004ApJ...611.1005G} that has been operational since 2004. It features an effective area of $110 \text{ cm}^2$ at 1.5~keV, a 23~arcmin field of view (FOV), and a spatial resolution of 18~arcsec (half-power diameter) \citep{2005SSRv..120..165B}. The X-ray light curves and spectra were generated using the \textit{Swift}-XRT online data product generator\footnote{\url{https://www.swift.ac.uk/user_objects/}} \citep{2007A&A...469..379E,2009MNRAS.397.1177E}. This pipeline provides high-level products including light curves, spectra, and enhanced positions for point sources \citep{2025MNRAS.544.2455N}. For the spectral analysis, the $0.3\text{--}10.0$~keV data were grouped to a minimum of 20 counts per energy channel using the \texttt{grppha} tool to ensure the validity of $\chi^2$ statistics. The re-binned spectra were modeled in \texttt{XSPEC}\footnote{\url{https://heasarc.gsfc.nasa.gov/xanadu/xspec/}} (version 12.14.1) using an absorbed power-law model (\texttt{tbabs * powerlaw}). The \texttt{tbabs} component accounts for photoelectric absorption along the line of sight. Following the broadband X-ray spectral analysis of BL Lacertae performed by \citet{2022MNRAS.509...52D} using \textit{NICER} and \textit{NuSTAR} observations, we fixed the neutral hydrogen column density at $N_\mathrm{H} = 2.59 \times 10^{21}~\mathrm{cm}^{-2}$.

\subsection{\textit{Swift}-UVOT}
We utilized simultaneous observations from the \textit{Swift}-UVOT onboard the \textit{Neil Gehrels Swift Observatory} across all six filters, covering the full optical and ultraviolet range: $V$ ($500\text{--}600$ nm), $B$ ($380\text{--}500$ nm), $U$ ($300\text{--}400$ nm), $UVW1$ ($220\text{--}400$ nm), $UVM2$ ($200\text{--}280$ nm), and $UVW2$ ($180\text{--}260$ nm). Data processing was performed using the \texttt{HEASoft} package (v6.34). Aperture photometry was conducted on the processed images using the \texttt{uvotsource} tool. Source counts were extracted from a circular region with a radius of $5^{\prime\prime}$ centered on the target coordinates. The background was estimated from a nearby source-free circular region with a radius of $20^{\prime\prime}$ \citep{2022MNRAS.509...52D}. The background-subtracted source counts were then used to derive the observed magnitudes and flux densities. To account for host-galaxy contamination, we performed a subtraction based on the flux densities reported by \citet{2013MNRAS.436.1530R}. Specifically, the host-galaxy contributions for the $V$, $B$, $U$, $UVW1$, $UVM2$, and $UVW2$ bands were taken as $2.89$, $1.30$, $0.36$, $0.026$, $0.020$, and $0.017$~mJy, respectively. These values, representing approximately 50\% of the total galaxy flux captured within the UVOT aperture, were subtracted from the observed photometry to obtain the net flux of the source. Galactic extinction correction was then applied using $E(B-V)=0.291$ \citep{2011ApJ...737..103S} and the mean extinction law from \citet{1989ApJ...345..245C}. The corrected magnitudes were subsequently converted into flux densities using the zero points and conversion factors provided by \citet{2008MNRAS.383..627P} and \citet{2008AIPC.1065...81R}.

\subsection{Mephisto}
From MJD 60600 to 60635 (17 October -- 21 November 2024), BL Lacertae was monitored for 23 nights using the Multi-channel Photometric Survey Telescope (Mephisto). Mephisto is a 1.6-m wide-field, multi-channel telescope operated by Yunnan University and located at the Lijiang Observatory in China. Mephisto utilizes a Ritchey-Chr{\'e}tien (RC) optical system equipped with correctors and film-coated cubic prisms. The telescope is outfitted with three-channel single-chip CCD cameras covering one-quarter of the field of view (FOV). By employing dichroic prisms, Mephisto enables simultaneous multi-band observations across three arms: $uv$, $gr$, and $iz$. To maintain a high signal-to-noise ratio (SNR) while ensuring intensive temporal sampling, exposure times between 10 and 100~s were adopted. During this campaign, we collected a total of 539 data points over 20 observing days across six optical bands ($u$, $v$, $g$, $r$, $i$, $z$): 41 in $u$, 41 in $v$, 166 in $g$, 160 in $r$, 67 in $i$, and 64 in $z$. The raw frames were processed using a custom pipeline developed for Mephisto, which performed bias and dark subtraction, flat-field correction, and cosmic ray removal; details of the data reduction are given in \citet{2026MNRAS.548ag501C}. After obtaining the differential photometric measurements, we further corrected the magnitudes for Galactic interstellar extinction. Adopting the extinction law of \citet{1989ApJ...345..245C} and $E(B-V) = 0.291$ \citep{2011ApJ...737..103S}, together with the mean wavelengths of the Mephisto filters \citep{2024ApJ...969..126Y}, the extinction values for the $u$, $v$, $g$, $r$, $i$, and $z$ bands were calculated to be 1.40, 1.31, 0.92, 0.76, 0.49, and 0.39 mag, respectively. The extinction-corrected magnitudes were then converted into flux densities.

\subsection{Zwicky Transient Facility}
We retrieved $g$- and $r$-band light curves for BL Lacertae from the Zwicky Transient Facility (ZTF; \citealt{2019PASP..131a8003M}) time-domain survey. The ZTF utilizes the 48-inch Samuel Oschin Schmidt telescope, which features a $48~\text{deg}^2$ field of view, to monitor the sky in $g$, $r$, and $i$ optical bands. With a typical exposure time of 30~s, the survey reaches a limiting magnitude of $\sim 20.5$~mag in the $r$-band \citep{2019PASP..131a8002B}. To ensure high data quality, we filtered the observations by requiring a quality score of \texttt{catflags} = 0, as specified in the ZTF documentation. For nights with multiple intra-night observations, we calculated the daily average of the data points to obtain a single flux measurement per day. Similarly, the Galactic extinction correction was applied to the ZTF $g$- and $r$-band data, with corresponding extinction values of 1.02 and 0.73 mag, respectively, calculated using the mean wavelengths of the ZTF filter system \citep{2024A&A...689A..93R}. The corrected magnitudes were subsequently converted into flux densities. Due to the lack of multiple observations per night from both ZTF and \textit{Swift}-UVOT during the periods of peak activity, an intra-night variability analysis could not be conducted using these datasets.

\section{Results}
\label{sec_analysis}
The multi-wavelength light curves of BL Lacertae are presented in Figure \ref{f_1}. The first panel displays the 1-day binned \textit{Fermi}-LAT light curve in the 0.1--300 GeV energy range from MJD 60500 to 60800 (9 July 2024 - 5 May 2025). Corresponding light curves for \textit{Swift}-XRT and UVOT are shown in the second and third panels, respectively, while the ZTF and Mephisto light curves are provided in the fourth and fifth panels. The \textit{Fermi}-LAT light curve covers the major activity period observed on 5 October 2024 (MJD 60588), as well as a secondary active state on 12 February 2025 (MJD 60718).

\subsection{Orbit-binned $\gamma$-ray Variability}
\label{sec_gamma}
\textit{Fermi}-LAT $\gamma$-ray light curves were generated by modeling the spectra in each time interval with a simple power-law (PL) model. Compared to the more complex log-parabola (LP) model, the PL index generally yields smaller statistical uncertainties \citep[e.g.,][]{2011ApJ...733L..26A}. To identify periods of activity within the 1-day binned $\gamma$-ray light curve, we employed the Bayesian Blocks (BB) algorithm \citep{2013ApJ...764..167S} with a false-positive probability of $p_0 = 0.05$. Subsequently, the HOP algorithm was used to characterize these flares, applying a threshold that required the Bayesian block flux to exceed three times the mean flux ($F_{BB} \geq 3\bar{F}$). This analysis was conducted using the Python package \texttt{lightcurves}\footnote{\url{https://github.com/swagner-astro/lightcurves}} \citep{2022icrc.confE.868W}. As shown in the top panel of Figure \ref{f_1}, two primary active states were identified: MJD 60583--60594 (30 September -- 11 October 2024; F1) and MJD 60716--60726 (10 February -- 20 February 2025; F2), both denoted by pink shaded regions.

We first generated light curves binned on the \textit{Fermi} orbital timescale (approximately 96 minutes). We estimated the shortest flux doubling and halving timescales during these intervals using the following relation:
\begin{equation}
F(t_2) = F(t_1) \times 2^{(t_2 - t_1) / \tau},
\label{eq:1}
\end{equation}
where $F(t_1)$ and $F(t_2)$ are the fluxes at times $t_1$ and $t_2$, respectively, and $\tau$ is the flux doubling or halving timescale. For the first active state F1, the shortest significant flux doubling timescale was found to be $\tau = 1.47 \pm 0.30$ hr (at MJD 60587.20; 4 October 2024) with a significance of $6.0\sigma$\footnote{The significance is calculated as $\sigma = \frac{\vert{}F(t_2) - F(t_1)\vert{}}{\sqrt{\sigma_{t1}^2 + \sigma_{t2}^2}}$, where $\sigma_{t1}$ and $\sigma_{t2}$ are the corresponding uncertainties of the fluxes \citep{2022MNRAS.509...52D}.}, while for the second active state F2, the shortest significant halving timescale was $\tau = 1.33 \pm 0.29$ hr (at MJD 60718.40; 12 February 2025) with a significance of $5.9\sigma$.

\begin{figure*}
\centering
\includegraphics[scale=.20]{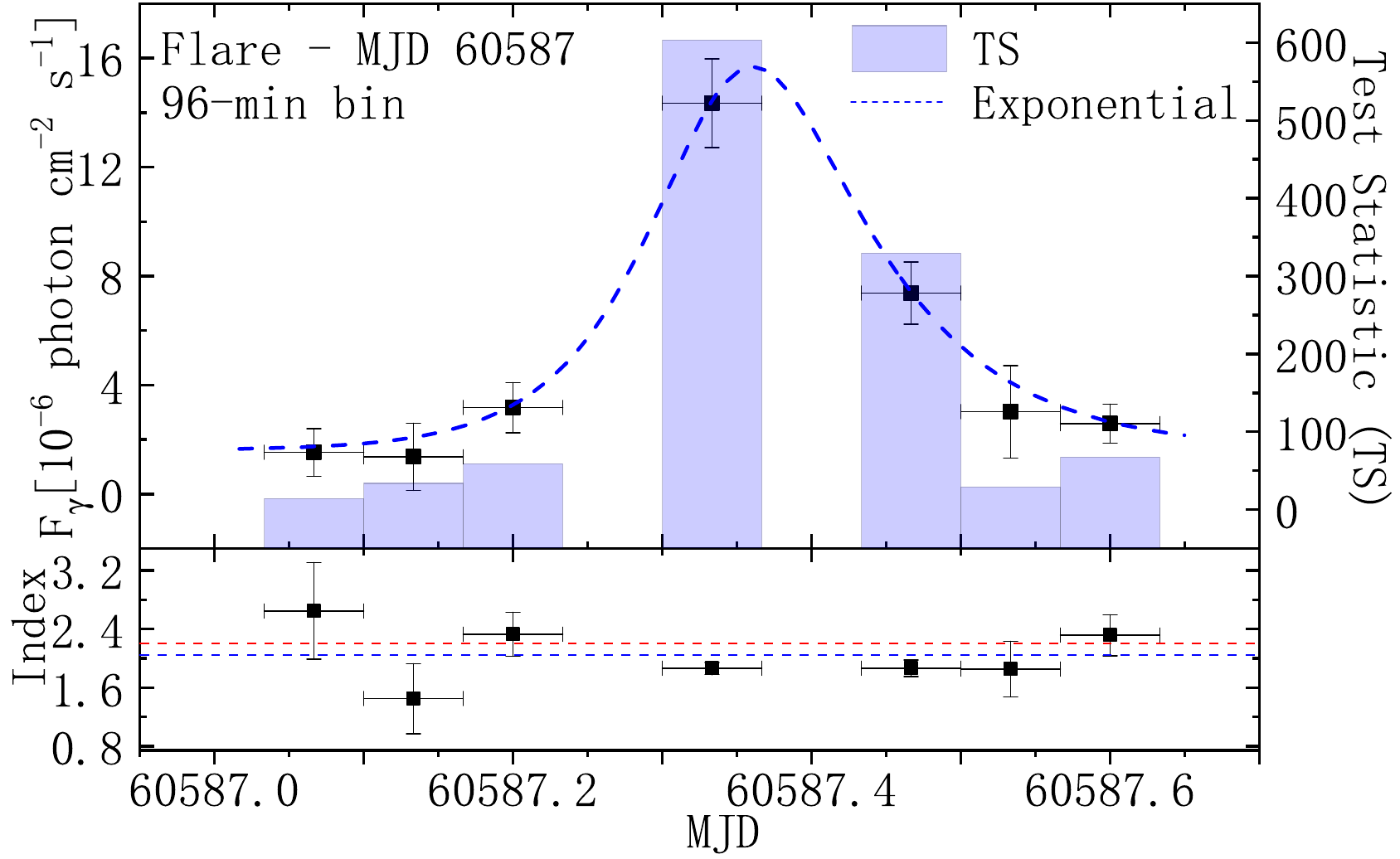}
\includegraphics[scale=.20]{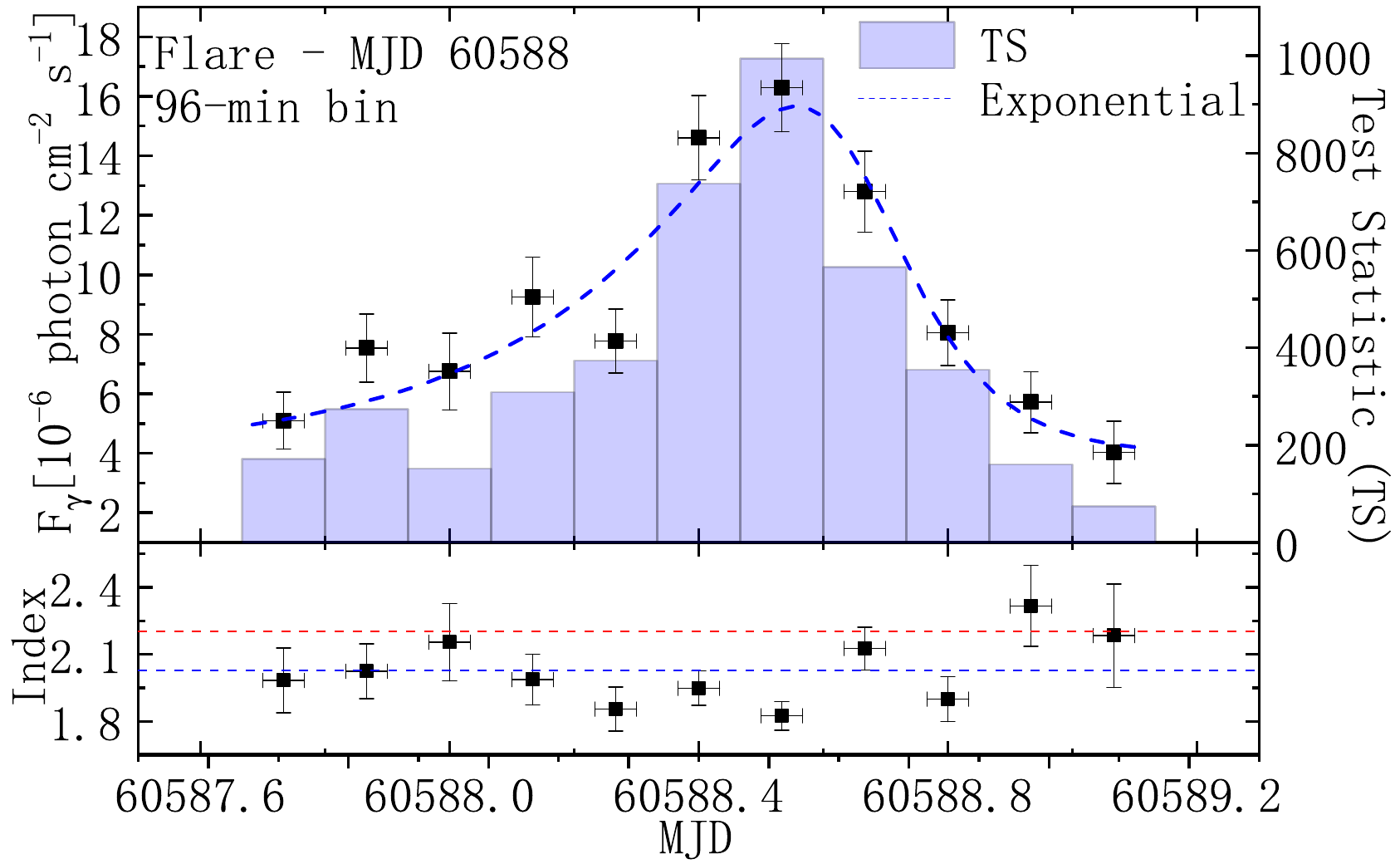}
\includegraphics[scale=.20]{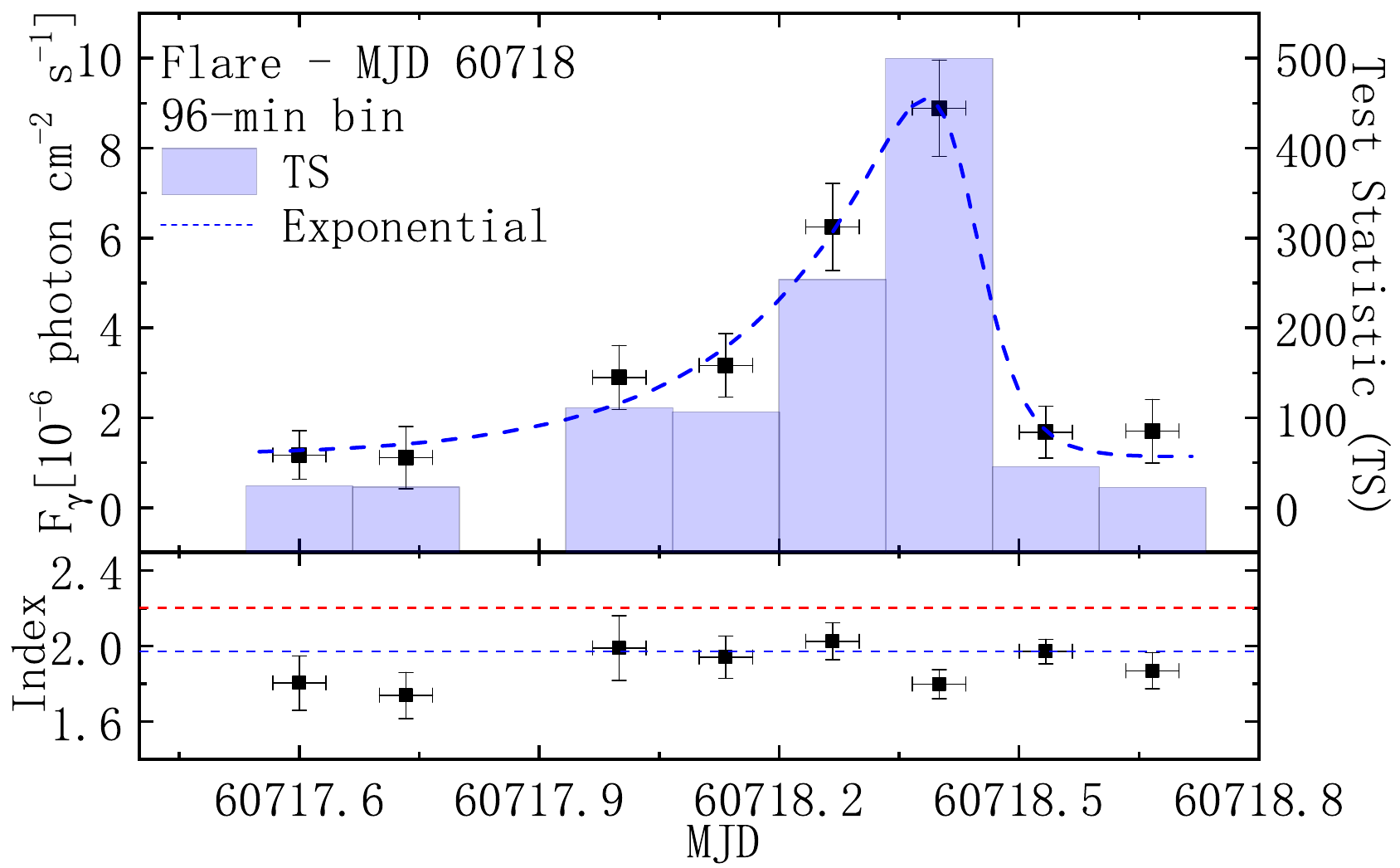}
\caption{Orbit-binned $\gamma$-ray light curves of BL Lacertae for MJD 60587, MJD 60588, and MJD 60718 (4 October 2024, 5 October 2024, and 12 February 2025). The TS values for each time bin are represented by blue histograms, while the blue dashed lines indicate the exponential fitting results. The bottom panels illustrate the PL photon indices for each corresponding bin. In these panels, the red and blue horizontal lines represent the 4FGL catalog PL index ($2.20$) and the daily average photon index, respectively. The data behind this figure are available in machine-readable format in the online journal.}
\label{f_2}
\end{figure*}

To characterize the temporal profiles of individual flares, the orbit-binned light curves were modeled using a double-exponential flare profile following \citet{2010ApJ...722..520A}, a functional form well suited for evaluating both flare duration and symmetry:
\begin{equation}
F(t) = F_{\mathrm{c}} + F_0 \left( e^{\frac{t_0 - t}{T_{\mathrm{r}}}} + e^{\frac{t - t_0}{T_{\mathrm{d}}}} \right)^{-1},
\label{eq:2}
\end{equation}
where $F_{\mathrm{c}}$ represents an assumed constant baseline level underlying the flare, $F_0$ and $t_0$ represent the flare amplitude and peak time, while $T_{\mathrm{r}}$ and $T_{\mathrm{d}}$ denote the rise and decay timescales, respectively. Exponential fitting was performed for the major outburst at MJD 60588 (5 October 2024; Flare~II) and its preceding sub-flare at MJD 60587 (4 October 2024; Flare~I) within the F1 interval, as well as for the outburst at MJD 60718 (12 February 2025; Flare~III) within the F2 interval. Figure \ref{f_2} displays the $\gamma$-ray light curves and the corresponding best-fit profiles for these three flares. The peaks of Flares~I, II, and III were identified at MJD $60587.35 \pm 0.02$, $60588.64 \pm 0.06$, and $60718.43 \pm 0.03$, respectively. We estimated the asymmetry parameter, $\xi = (T_{\mathrm{d}} - T_{\mathrm{r}}) / (T_{\mathrm{d}} + T_{\mathrm{r}})$, to characterize the flare symmetry \citep{2010ApJ...722..520A}. The results indicate that Flare~I exhibits a nearly symmetric structure ($\xi = 0.2$), whereas Flare~II and Flare~III display moderate asymmetry, with $\xi$ values of $-0.5$ and $-0.7$, respectively. The parameters derived from the exponential fits are summarized in Table \ref{table1}. We further investigated the spectral variability during these flaring episodes. The average photon indices of Flares~I, II, and III are $2.04 \pm 0.32$, $2.02 \pm 0.12$, and $1.89 \pm 0.11$, respectively. These photon indices are generally harder than the photon index reported in the 4FGL catalog \citep{2020ApJS..247...33A} ($\Gamma = 2.20$). A Spearman's rank correlation analysis performed on the orbit-binned light curves during these flaring episodes reveals no significant correlation between the $\gamma$-ray flux and photon index on this timescale.

\begin{deluxetable*}{ccccccccc}
\renewcommand\arraystretch{1}
\setlength{\tabcolsep}{3.8mm}
\centering
\tablecaption{Results of the exponential fit to the orbit-binned lightcurves.}
\tablehead{Flare & Gregorian Date & $T_{\rm peak}$ (MJD) & $F_{\rm peak}$ ($10^{-6}$ ph cm$^{-2}$ s$^{-1}$) & $t_{\rm r}$ (hr) & $t_{\rm d}$ (hr) & $\xi$}
\startdata
Flare~I	& 4 October 2024 &	60587.35$\pm$0.02 	&	15.4$\pm$1.3 	&	1.3$\pm$0.1 	&	1.8$\pm$0.3	&	0.2\\			
Flare~II	& 5 October 2024 &	60588.64$\pm$0.06 	&	14.2$\pm$1.6 	&	7.7$\pm$2.0 	&	2.5$\pm$0.9	&	-0.5\\
Flare~III	& 12 February 2025 &	60718.43$\pm$0.03 	&	9.1$\pm$0.8 	&	4.5$\pm$1.0 	&	0.8$\pm$0.2	&	-0.7\\
\enddata
\tablecomments{
Column 1: Number of the flares;\\
Column 2: Gregorian Date;\\
Column 3: Peak time ($T_{\rm peak}$) in MJD;\\
Column 4: Peak flux of the 96-min bin light curve;\\
Column 5: Rise time ($t_{\rm r}$) in hours;\\
Column 6: Decay time ($t_{\rm d}$) in hours;\\
Column 7: Flare asymmetry parameter ($\xi$).\\
}
\label{table1}
\end{deluxetable*}

\subsection{Sub-minute $\gamma$-ray Fast Variability}
\label{sec_subminute}
The orbit-binned (96-minute) light curves reveal three prominent flux peaks with high photon statistics at MJD 60587.33 (4 October 2024) ($F = (14.4 \pm 1.6) \times 10^{-6} \, \mathrm{ph\,cm^{-2}\,s^{-1}}$; $\mathrm{TS} = 556$), MJD 60588.53 (5 October 2024) ($F = (16.3 \pm 1.5) \times 10^{-6} \, \mathrm{ph\,cm^{-2}\,s^{-1}}$; $\mathrm{TS} = 994$), and MJD 60718.40 (12 February 2025) ($F = (8.9 \pm 1.1) \times 10^{-6} \, \mathrm{ph\,cm^{-2}\,s^{-1}}$; $\mathrm{TS} = 500$). The high statistical significance of these orbital peaks enables the investigation of ultra-fast $\gamma$-ray variations on timescales of a few minutes. For the minute-scale variability analysis, we used spacecraft location and attitude data from the 1-second resolution FT2 files to account for temporal variations in the effective area and inclination angle.
Subsequently, we generated 2-, 3-, and 5-minute binned light curves to search for such rapid variability. Following \citet{2016ApJ...824L..20A} and \citet{2018ApJ...854L..26S}, we tested for minute-scale variability by fitting a constant flux model to each orbit. We then calculated the $p$-value under the null hypothesis of constant flux. We adopted a threshold of $p \le 0.05$ (a 95\% confidence level) to identify potential sub-orbital variability \citep{2022A&A...668A.152P}. We note that this criterion is intended as an indication of possible variability rather than a stringent detection threshold. If a more stringent threshold (e.g., $p \le 0.01$) were applied, the variability reported here would not satisfy the stricter requirement and should therefore be regarded as tentative evidence rather than a statistically significant detection.

\begin{figure*}
\centering
\includegraphics[scale=.32]{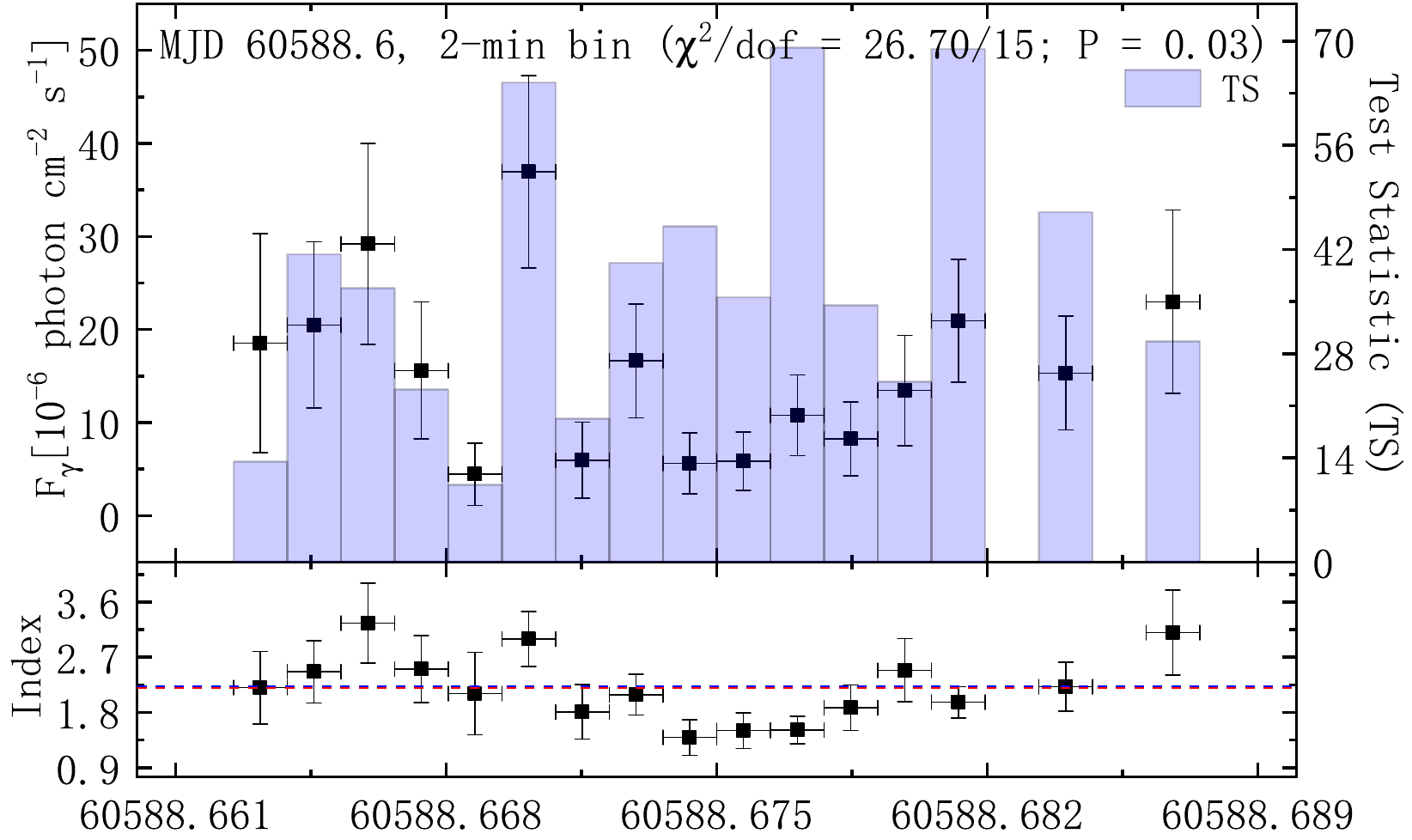}
\caption{2-minute binned $\gamma$-ray light curve of BL Lacertae during the main flare around MJD 60588.6 (5 October 2024). The upper panel shows the TS values as blue histograms. The bottom panel displays the corresponding PL photon indices. The red and blue horizontal lines in the bottom panel denote the 4FGL catalog PL index ($2.20$) and the average photon index, respectively. The data behind this figure are available in machine-readable format in the online journal.}
\label{f_3}
\end{figure*}

Using the 1-second resolution data, no significant minute-scale flux variability was detected during the Flares~I and III, where the $p$-values for all time bin sizes were consistent with a constant flux. However, tentative evidence for minute-scale variability was found in the 2-minute binned light curves during the major outburst around MJD 60588.6 (5 October 2024; Flares~II). The statistical result for this epoch yields $p = 0.03$ ($\chi^2/\text{dof} = 26.70/15$) for MJD 60588.67, which satisfies our $p \le 0.05$ criterion but represents marginal evidence rather than a significant detection. The corresponding shortest variability timescales were estimated using Eq. (\ref{eq:1}) as $\tau = 0.7 \pm 0.2 \, \mathrm{min}$ for flux doubling. For the 3- and 5-minute binned data, the $p$-values increased above the significance threshold. The reduction in significance is expected, as larger time bins tend to smooth out fast variations during bright flares \citep{2012A&A...544A...6L}. Conversely, shorter time bins can lead to increased statistical uncertainties and lower signal-to-noise ratios due to limited photon counts \citep{2012A&A...544A...6L}. Notably, our derived timescales are shorter than the $\sim 0.9 \pm 0.4 \, \mathrm{min}$ timescale previously detected in BL Lacertae by \citet{2022A&A...668A.152P}. More extensive observations are required to further refine these findings. The 2-minute binned light curves are presented in Figure \ref{f_3}.

\subsection{The $\gamma$-ray Spectral Variability}
\begin{figure*}
\centering
\includegraphics[scale=.27]{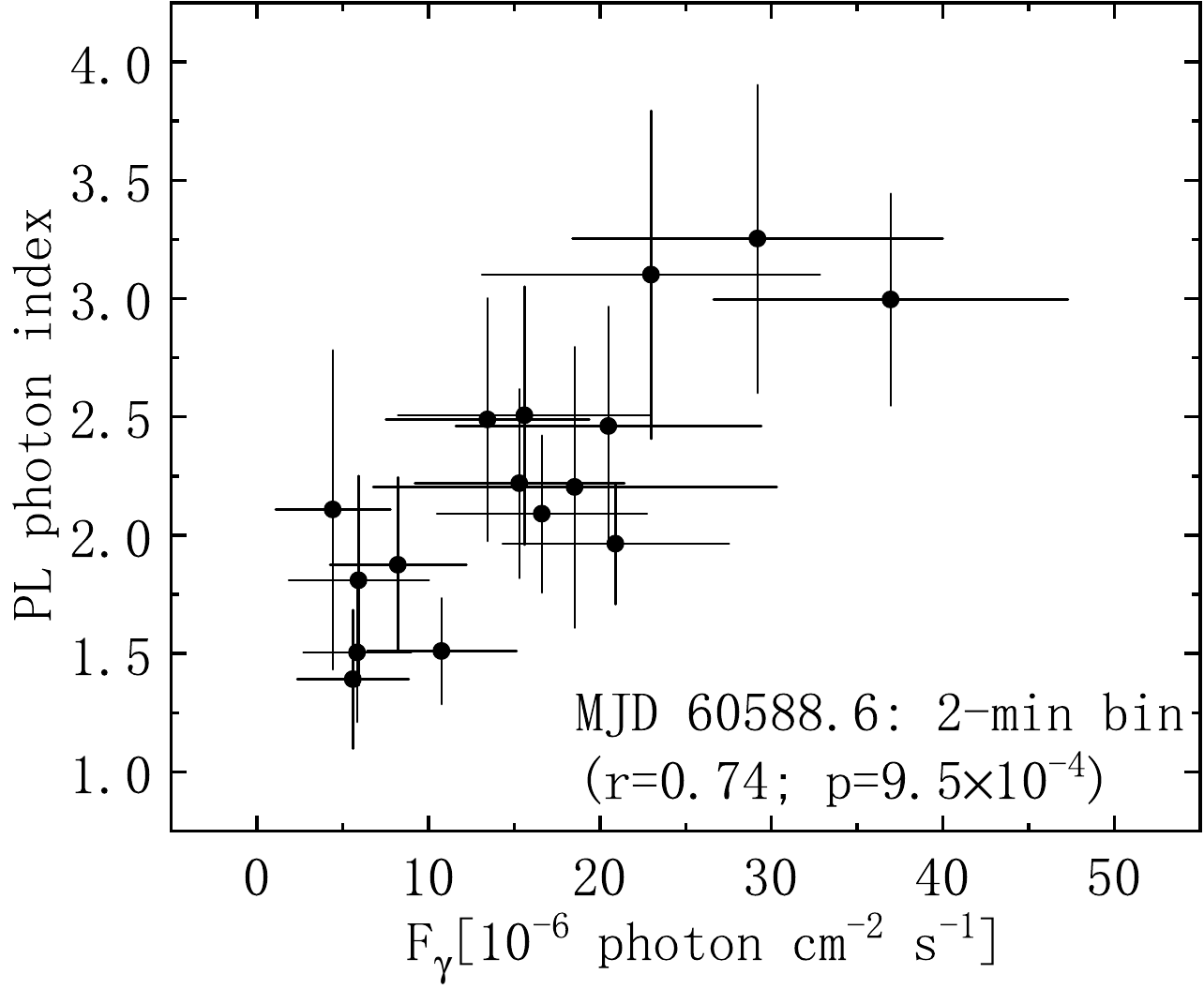}
\includegraphics[scale=.27]{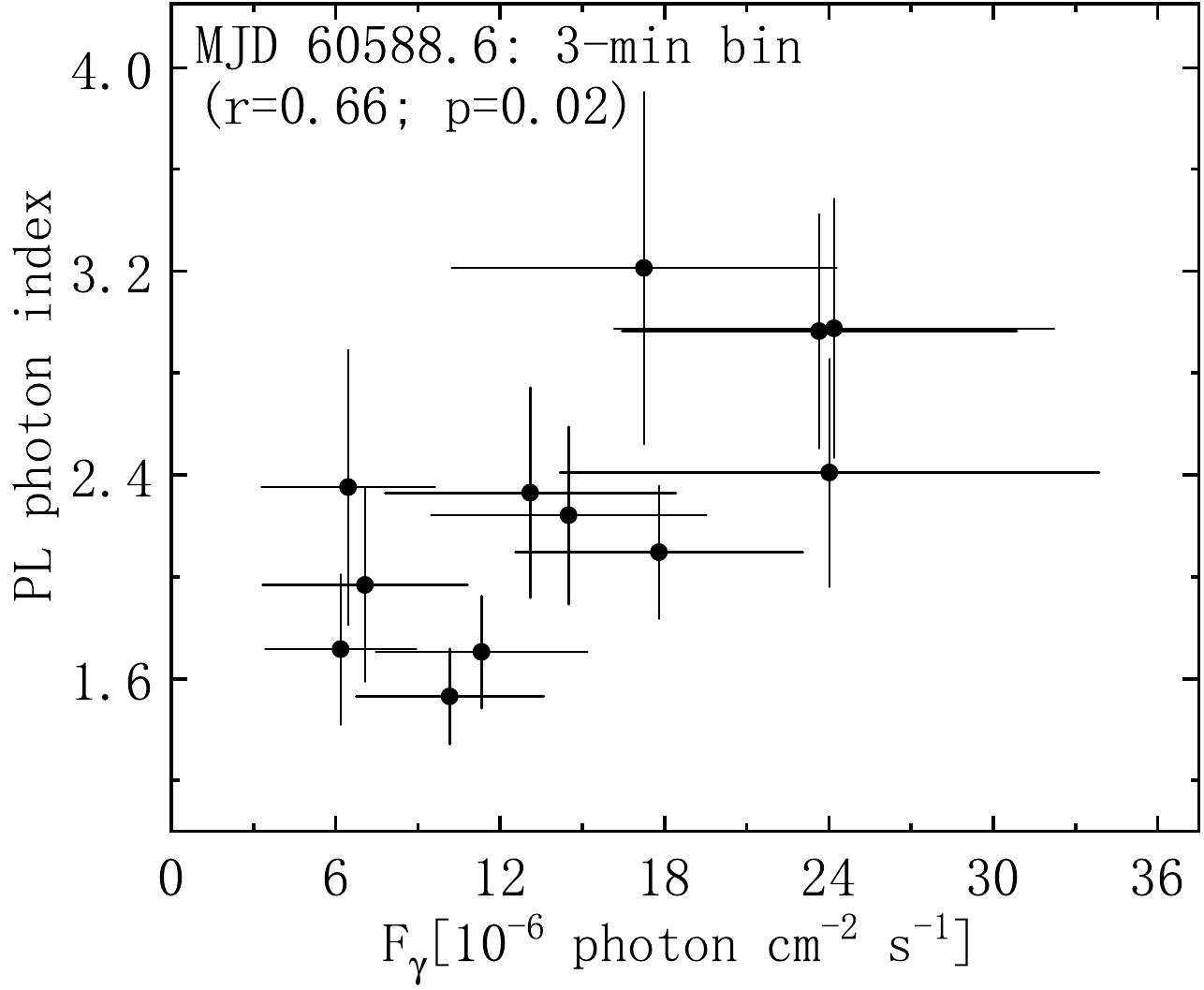}
\includegraphics[scale=.27]{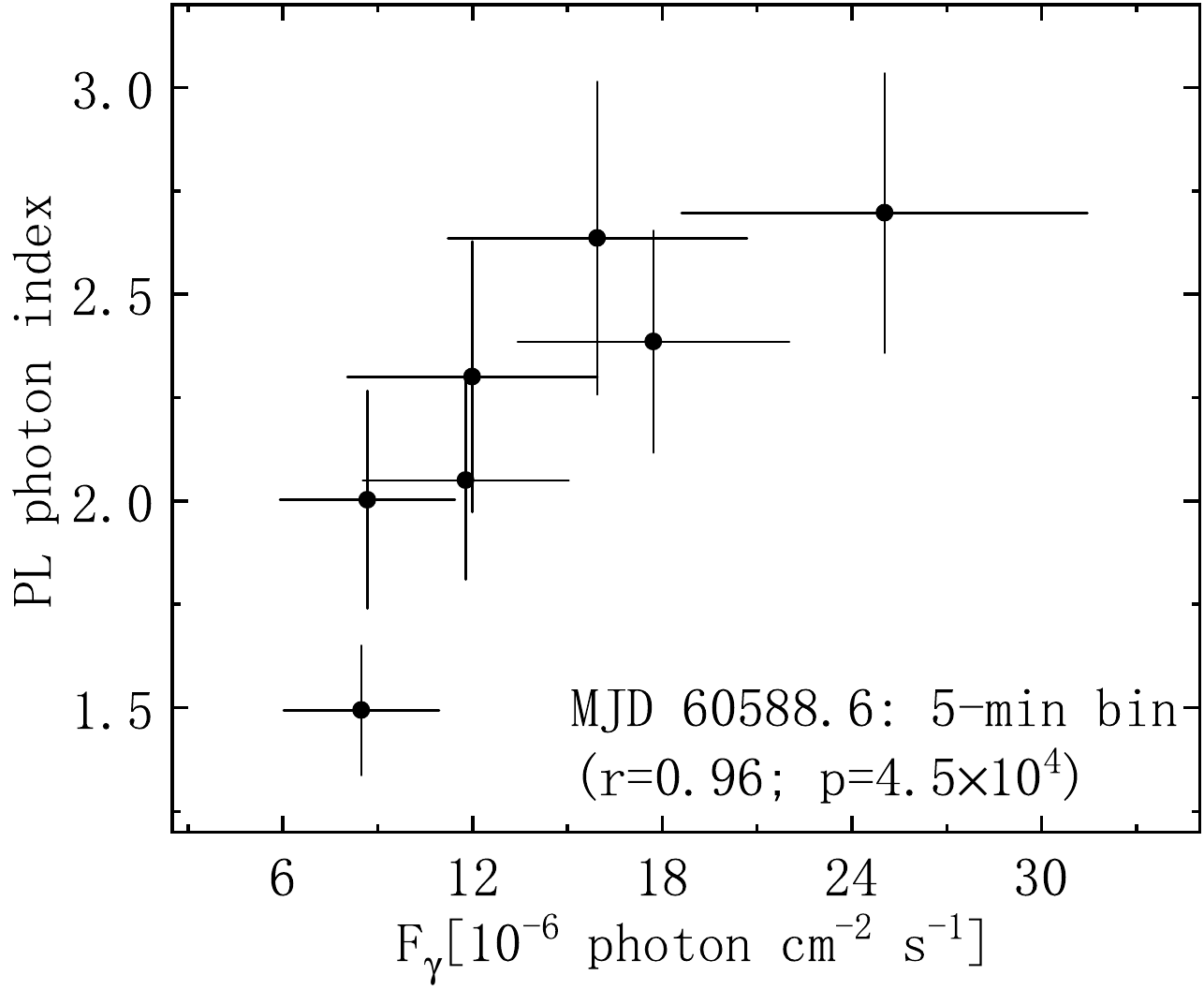}
\caption{Variations of the PL photon index with $\gamma$-ray flux during the main flare on MJD 60588 (5 October 2024). The left, middle, and right panels show the correlation between the PL photon index and $\gamma$-ray flux for the 2-, 3-, and 5-minute binned light curves, respectively. $r$ is the Spearman rank correlation coefficient, and $p$ is the corresponding chance probability. A clear ``softer-when-brighter'' trend is evident in all three binning intervals. The data behind this figure are available in machine-readable format in the online journal.}
\label{f_4}
\end{figure*}

We investigated the $\gamma$-ray spectral variability of BL Lacertae during MJD 60587, 60588, and 60718 (4 October 2024, 5 October 2024, and 12 February 2025) using light curves binned on daily, orbital, and short-term timescales. Across the daily and orbital binning intervals, the average PL photon indices were consistently harder than the 4FGL catalog value of 2.20. In the 2-, 3-, and 5-minute binning intervals (based on 1-second resolution FT2 data), however, the average PL photon indices were softer than 2.20 only at MJD 60588.6 ($2.22 \pm 0.45$, $2.29 \pm 0.39$ and $2.22 \pm 0.28$, respectively). To quantify the spectral evolution, we examined the relationship between the observed $\gamma$-ray flux and the PL photon index using Spearman's rank correlation analysis. Significant positive correlations were identified in the 2- and 5-minute binned light curves around MJD 60588.6, with correlation coefficients of $r = 0.74$ and $r = 0.96$, and null hypothesis probabilities of $p = 9.5 \times 10^{-4}$ and $p = 4.5 \times 10^{-4}$, respectively. A weaker positive correlation was also observed in the 3-minute binned light curve ($r = 0.66$, $p = 0.02$). Table~\ref{table2} presents the mean PL photon indices and correlation results. Figure~\ref{f_4} illustrates the relationship between the $\gamma$-ray flux and the photon index, revealing a characteristic ``softer-when-brighter'' behavior.

\begin{deluxetable*}{ccccccccc}
\renewcommand\arraystretch{1}
\setlength{\tabcolsep}{5mm}
\centering
\tablecaption{Results of spectral variability of BL Lacertae in different time bins.}
\tablehead{Bin\ size  &  Gregorian Date  &  Time (MJD)   &   Average\ PL\ index  &  $r$   &  $p$}
\startdata
2-min	& 5 October 2024  &	60588.6--60588.7   &   2.22$\pm$0.45 	&	0.74	&	$9.5 \times 10^{-4}$ \\			
3-min	& 5 October 2024  &	60588.6--60588.7   &   2.29$\pm$0.39 	&	0.66	&	0.02 \\	
5-min	& 5 October 2024  &	60588.6--60588.7   &   2.22$\pm$0.28 	&	0.96	&	$4.5 \times 10^{-4}$ \\	
\enddata
\tablecomments{
Column 1: Bin size of the light curves;\\
Column 2: Gregorian Date;\\
Column 3: The time range of Flux vs PL index;\\
Column 4: The average PL index;\\
Column 5: The Spearman's rank correlation coefficient;\\
Column 6: The null hypothesis probability.\\
}
\label{table2}
\end{deluxetable*}

\subsection{Broadband SED Modeling}
Quasi-simultaneous broadband SEDs are constructed using multi-wavelength data from \textit{Fermi}-LAT, \textit{Swift}-XRT/UVOT, ZTF, and Mephisto for the two flaring epochs, F1 and F2. The F1 SED corresponds to MJD~60588 (5 October 2024), while the F2 SED is characterized using nearly simultaneous optical and X-ray observations on MJD~60719 (13 February 2025), together with \textit{Fermi}-LAT data averaged over MJD~60718--60719 (12 February 2025 -- 13 February 2025). To investigate the spectral evolution, we also construct a representative low-flux quiet-state SED using \textit{Swift}-XRT observations on MJD~60602, 60610, and 60623 (19 October 2024, 27 October 2024, and 9 November 2024), optical data obtained within one day of these X-ray epochs, and \textit{Fermi}-LAT data averaged over MJD~60597--60625 (14 October 2024 -- 11 November 2024). To facilitate the fitting process, the observational data are logarithmically binned with a bin width of 0.1 dex in frequency. We employ a one-zone leptonic model using the \texttt{JetSet} code \footnote{\url{https://jetset.readthedocs.io/en/1.3.0/index.html}} \citep{2020ascl.soft09001T} to fit the SEDs. Within this framework, the emission region is assumed to be a homogeneous sphere of radius $R$. The underlying electron distribution, $N(\gamma)$, is modeled as a broken power law \citep{2010MNRAS.401.1570T}, defined as:
\begin{equation}
N(\gamma) = N_e \begin{cases} \gamma^{-p_1} & \gamma_{\min} \le \gamma \le \gamma_b \\ \gamma_b^{p_2-p_1} \gamma^{-p_2} & \gamma_b < \gamma \le \gamma_{\max} \end{cases}
\label{eq:3}
\end{equation}
where $\gamma_{\min}$, $\gamma_b$, and $\gamma_{\max}$ are the minimum, break, and maximum Lorentz factors, respectively; $p_1$ and $p_2$ denote the spectral indices below and above the break, respectively; and $N_e$ is the normalization constant in units of $\text{cm}^{-3}$.

The broadband jet emission is further governed by $R$, the magnetic field strength ($B$), the bulk Lorentz factor ($\Gamma$), and the viewing angle ($\theta$). Given the evidence for the BLR and DT in BL Lacertae across various epochs \citep{2021MNRAS.507.5602P}, we incorporate these external radiation fields into our SED modeling. Following the method of \citet{1997MNRAS.286..415C}, \citet{2009MNRAS.396..984G} adopted the line ratios from \citet{1991ApJ...373..465F} and included the contribution of the H$\alpha$ line to estimate the BLR luminosity of BL Lacertae as $L_{\mathrm{BLR}} \approx 10^{42.38} \,\mathrm{erg\,s^{-1}}$. Assuming $L_{\mathrm{disk}} \sim 10 L_{\mathrm{BLR}}$ \citep[e.g.,][]{2011MNRAS.414.2674G}, we obtain an accretion disk luminosity of $L_{\mathrm{disk}} \approx 2.4 \times 10^{43} \,\mathrm{erg\,s^{-1}}$. The characteristic radii of the BLR are determined according to the scaling relations of \citet{2007ApJ...659..997K}:
\begin{equation}
R_{\text{BLR-in}} = 10^{17} \times \sqrt{\frac{L_{\text{disk}}}{10^{45}}} \text{ cm},
\label{eq:4}
\end{equation}
\begin{equation}
R_{\text{BLR-out}} = 1.1 \times R_{\text{BLR-in}},
\label{eq:5}
\end{equation}
The radius of the dusty torus ($R_{\rm DT}$) depends on the accretion disk luminosity, following the expression given by \citet{2008MNRAS.387.1669G}:
\begin{equation}
R_{\text{DT}} = 2.5 \times 10^{18} \times \sqrt{\frac{L_{\text{disk}}}{10^{45}}} \text{ cm},
\label{eq:6}
\end{equation}

\begin{figure*}
\centering
\includegraphics[scale=.28]{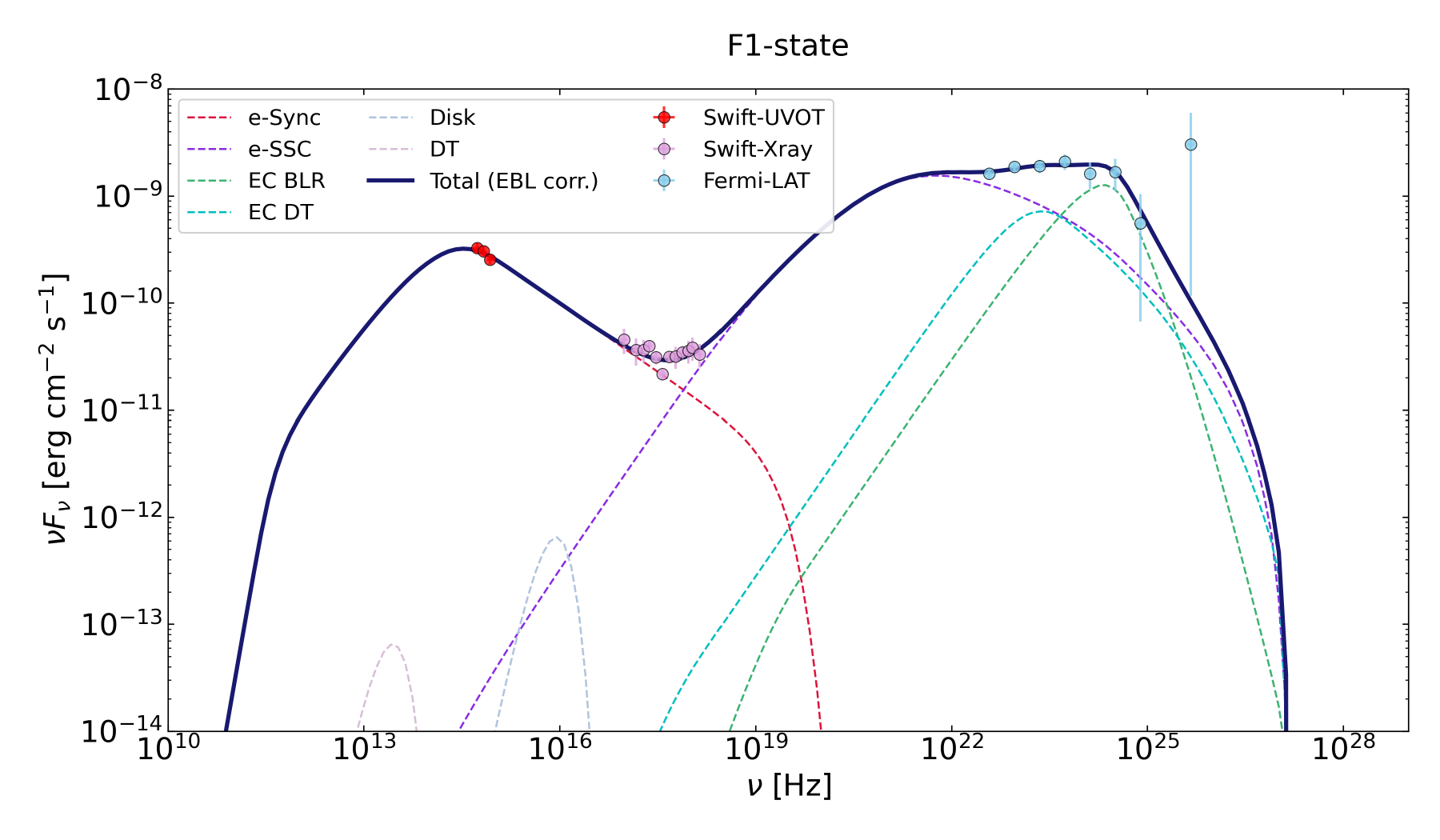}
\includegraphics[scale=.28]{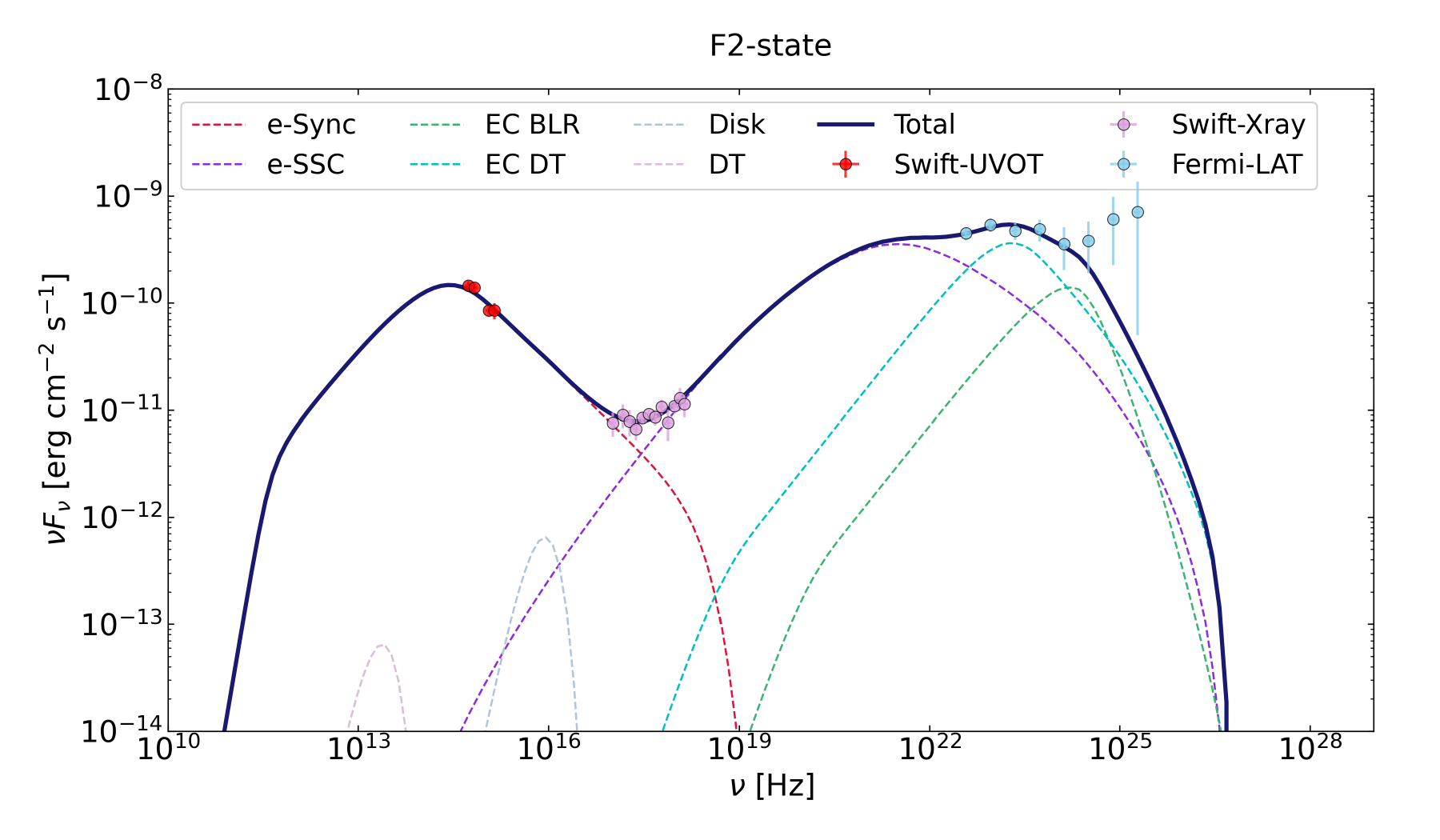}
\includegraphics[scale=.28]{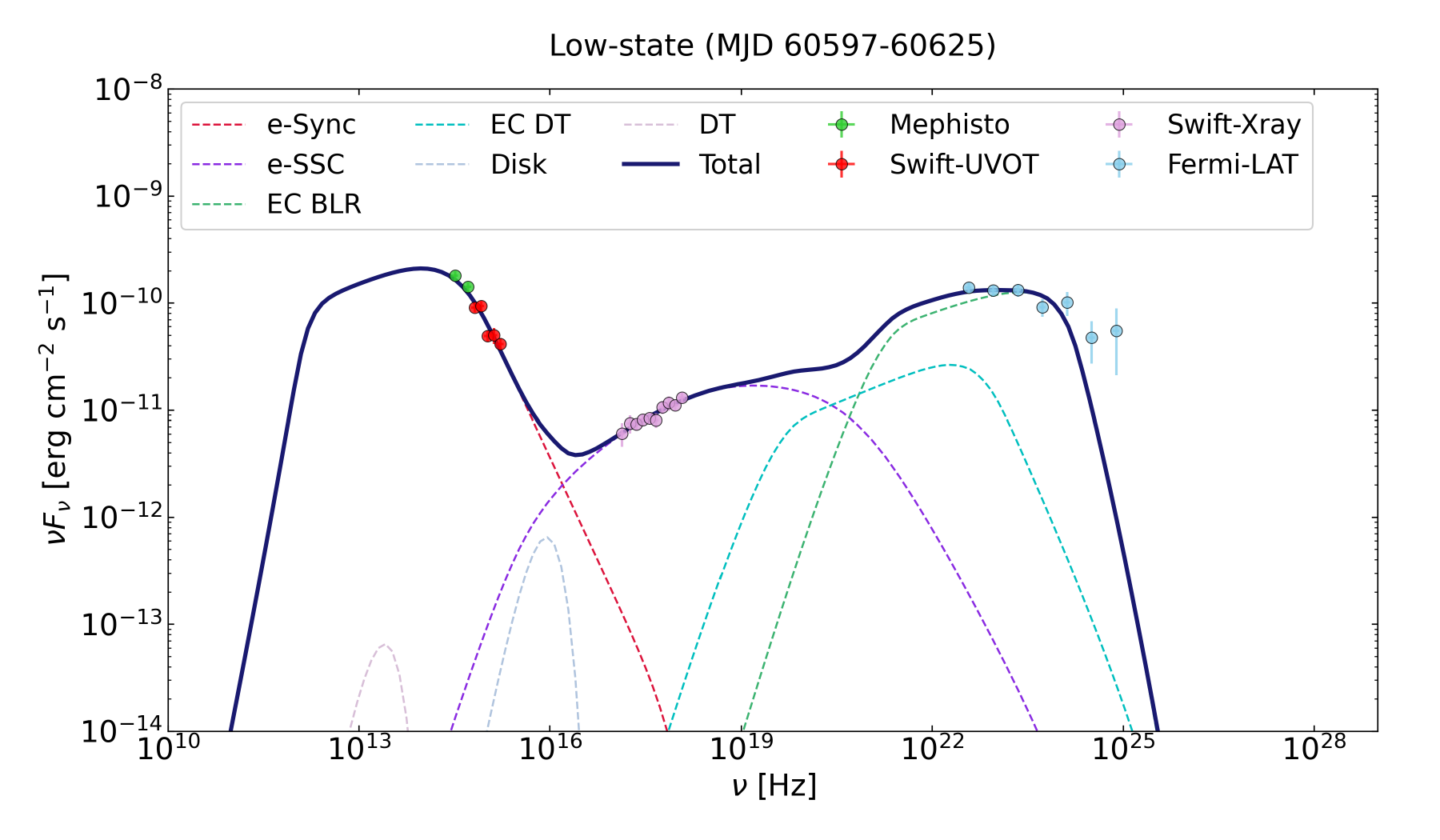}
\caption{Quasi-simultaneous broadband SEDs of BL Lacertae during the flaring epochs F1 (MJD~60588 (5 October 2024); top left panel) and F2 (MJD~60719 (13 February 2025), with \textit{Fermi}-LAT data averaged over MJD~60718--60719 (12 February 2025 -- 13 February 2025); top right panel), and the low-flux quiet state (bottom panel; based on X-ray/optical observations on MJD~60602, 60610, and 60623 (19 October 2024, 27 October 2024, and 9 November 2024), and \textit{Fermi}-LAT data averaged over MJD~60597--60625 (14 October 2024 -- 11 November 2024). Solid lines represent the best-fit results derived from a one-zone leptonic model.}
\label{f_5}
\end{figure*}

We adopt a viewing angle of $\theta = 0.1^\circ$ as a representative value, following previous modeling studies of BL Lacertae \citep{2025JHEAp..4800402M,2024MNRAS.527.5140S}, and assume that each relativistic electron is accompanied by one cold proton \citep{2024ApJ...974..233K}. The temperatures of the DT ($T_{\rm DT}$), BLR ($T_{\rm BLR}$), and accretion disk ($T_{\rm Disk}$) are fixed \citep{2025JHEAp..4800402M,2021MNRAS.507.5602P}. The remaining parameters, including $N_e$, $p_{1}$, $p_{2}$, $R$, and the location of the emission region ($R_{\rm H}$), are kept free during the fitting. The corresponding results and best-fit model parameters are presented in Figure \ref{f_5} and Table \ref{table3}, respectively.

\begin{deluxetable*}{lcccc}
\renewcommand\arraystretch{1.2}
\tabletypesize{\scriptsize}
\centering
\tablecaption{SED Modeling Parameters of Low and Flare States}
\tablewidth{0pt}
\tablehead{
\colhead{Parameter} & \colhead{Symbol [Unit]} & \colhead{Low State} & \colhead{F1 State} & \colhead{F2 State}}
\startdata
\tableline
\multicolumn{5}{c}{\textbf{Fixed Parameters}} \\
\tableline
Viewing angle                    & $\theta$ [$^{\circ}$]                     & \multicolumn{3}{c}{0.1} \\
BLR inner radius                 & $R_{\text{BLR,in}}$ [cm]                  & \multicolumn{3}{c}{$1.5 \times 10^{16}$} \\
BLR outer radius                 & $R_{\text{BLR,out}}$ [cm]                 & \multicolumn{3}{c}{$1.7 \times 10^{16}$} \\
Disk luminosity                  & $L_{\text{disk}}$ [erg s$^{-1}$]          & \multicolumn{3}{c}{$2.4 \times 10^{43}$} \\
Dust torus radius                & $R_{\text{DT}}$ [cm]                      & \multicolumn{3}{c}{$3.9 \times 10^{17}$} \\
Dust torus temperature           & $T_{\text{DT}}$ [K]                       & \multicolumn{3}{c}{$3.3 \times 10^{2}$} \\
Disk temperature                 & $T_{\text{Disk}}$ [K]                     & \multicolumn{3}{c}{$1.2 \times 10^{5}$} \\
\tableline
\multicolumn{5}{c}{\textbf{Free Parameters}} \\
\tableline
Electron minimum Lorentz factor   & $\gamma_{e,\min}$                         & $24.7_{-2.2}^{+2.5}$                  & $3.3_{-0.7}^{+0.6}$                  & $11.7_{-1.5}^{+2.8}$ \\
Electron maximum Lorentz factor   & $\gamma_{e,\max}$                         & $3.7_{-0.3}^{+0.5} \times 10^{4}$   & $5.3_{-0.5}^{+0.6} \times 10^{5}$   & $1.9_{-0.4}^{+0.4} \times 10^{5}$ \\
Electron normalization              & $N_e$ [cm$^{-3}$]                         & $9.0_{-1.2}^{+0.8} \times 10^{3}$   & $1.3_{-0.1}^{+0.1} \times 10^{4}$   & $1.0_{-0.1}^{+0.1} \times 10^{4}$ \\
Electron break Lorentz factor    & $\gamma_{e,\text{break}}$                 & $8.9_{-0.8}^{+1.1} \times 10^{2}$   & $2.1_{-0.2}^{+0.2} \times 10^{3}$   & $2.3_{-0.1}^{+0.1} \times 10^{3}$ \\
Low energy electron spectral index    & $p_{1}$                                   & $2.5_{-0.1}^{+0.2}$                  & $1.2_{-0.1}^{+0.1}$                  & $1.5_{-0.1}^{+0.1}$ \\
High energy electron spectral index   & $p_{2}$                                   & $5.6_{-0.4}^{+0.5}$                  & $3.9_{-0.1}^{+0.1}$                  & $4.2_{-0.1}^{+0.1}$ \\
Emission region radius               & $R$ [cm]                                  & $1.7_{-0.1}^{+0.2} \times 10^{15}$   & $1.5_{-0.2}^{+0.1} \times 10^{15}$   & $1.7_{-0.2}^{+0.2} \times 10^{15}$ \\
Emission region distance         & $R_H$ [cm]                                & $2.3_{-0.1}^{+0.1} \times 10^{16}$   & $2.4_{-0.1}^{+0.1} \times 10^{16}$   & $3.3_{-0.3}^{+0.4} \times 10^{16}$ \\
Magnetic field                    & $B$ [\text{G}]                            & $2.3_{-0.2}^{+0.2}$                  & $0.5_{-0.04}^{+0.07}$                & $0.4_{-0.02}^{+0.04}$ \\
Bulk Lorentz factor               & $\Gamma$                                  & $17.1_{-1.1}^{+1.3}$                 & $14.9_{-0.9}^{+0.8}$                 & $14.8_{-0.7}^{+0.2}$ \\
Reduced Chi-Squared              & $\chi^2/\text{dof}$                        & $1.6$                                 & $1.1$                                 & $0.9$ \\
\tableline
\multicolumn{5}{c}{\textbf{Jet Luminosities}} \\
\tableline
Jet power in electrons           & $L_e$ [erg s$^{-1}$]                      & $3.7_{-0.3}^{+0.4} \times 10^{43}$  & $1.4_{-0.2}^{+0.1} \times 10^{44}$  & $9.5_{-0.7}^{+0.8} \times 10^{43}$ \\
Jet power in magnetic field      & $L_B$ [erg s$^{-1}$]                      & $1.7_{-0.3}^{+0.5} \times 10^{43}$  & $4.7_{-0.6}^{+1.0} \times 10^{41}$  & $4.2_{-1.0}^{+1.7} \times 10^{41}$ \\
Jet power in protons              & $L_p$ [erg s$^{-1}$]                      & $1.1_{-0.1}^{+0.3} \times 10^{45}$  & $9.8_{-3.0}^{+3.2} \times 10^{44}$  & $8.7_{-2.3}^{+3.0} \times 10^{44}$ \\
Total jet power                  & $L_{\text{jet}}$ [erg s$^{-1}$]           & $1.1_{-0.2}^{+0.3} \times 10^{45}$  & $1.1_{-0.3}^{+0.3} \times 10^{45}$  & $9.7_{-2.2}^{+2.9} \times 10^{44}$ \\
\enddata
\tablecomments{
The table compares the physical properties of the source during different activity states (Low state, Flare 1, and Flare 2);\\
Column 1: Physical parameters of the SED modeling;\\
Column 2: Symbols and corresponding units for each parameter;\\
Column 3: Best-fit values obtained during the ``Low State'' (quiet period) of the source;\\
Columns 4 \& 5: Best-fit values obtained during the two ``Flare States'' (F1 and F2 active periods) of the source.
}
\label{table3}
\end{deluxetable*}

\section{Discussion}
\leavevmode\label{sec_discuss}
\subsection{The $\gamma$-ray Intraday Variability}
\label{sec_int}
In the 1-day binned \textit{Fermi}-LAT light curve, the maximum daily-averaged $\gamma$-ray flux was recorded on MJD 60588 (5 October 2024), reaching $(1.03 \pm 0.05) \times 10^{-5} \text{ ph } \text{cm}^{-2} \text{ s}^{-1}$ (TS = 3792), the largest flare ever detected from this source. The corresponding photon index was $1.93 \pm 0.04$, slightly harder than the 4FGL value of 2.20. Two secondary flares occurred on MJD 60587 and MJD 60718 (4 October 2024 and 12 February 2025), with daily-averaged fluxes of $(5.96 \pm 0.41) \times 10^{-6} \text{ ph cm}^{-2} \text{ s}^{-1}$ (TS = 1538) and $(3.84 \pm 0.32) \times 10^{-6} \text{ ph cm}^{-2} \text{ s}^{-1}$ (TS = 1004), respectively. Their photon indices ($1.94 \pm 0.05$ and $1.87 \pm 0.07$) were both harder than 2.20. Subsequently, based on the results of the Bayesian analysis, we constructed $\gamma$-ray light curves with 96-minute (orbital), 5-minute, 3-minute, and 2-minute temporal bins.

The diversity of flare profiles reflects differences in the underlying emission mechanisms \citep{2010ApJ...722..520A}. Based on the orbital binned light curves, we calculated the asymmetry parameter, $\xi$, to characterize the flare symmetry. According to the classification criteria proposed by \citet{2010ApJ...722..520A}, the flare at MJD 60587 (4 October 2024) ($\xi = 0.2$) is classified as a symmetric flare ($-0.3 < \xi < 0.3$). Similarly, \citet{2022A&A...668A.152P} reported two symmetric flares in BL Lacertae during the epochs of MJD 59327 and MJD 59331 (23 April 2021 and 27 April 2021). The formation of symmetric profiles is generally attributed either to the crossing time of radiation (or particles) through the emission region or to the superposition of multiple short-timescale flaring events \citep{2010ApJ...722..520A,2024MNRAS.527.5140S}. The asymmetry in flare profiles can be attributed to the intensification and weakening of particle acceleration processes \citep{2024MNRAS.527.5140S}. The flares at MJD 60588 (5 October 2024) ($\xi = -0.5$) and MJD 60718 (12 February 2025) ($\xi = -0.7$) are both categorized as moderately asymmetric ($-0.7 < \xi < -0.3$), exhibiting rise timescales that exceed their decay timescales. In such asymmetric profiles, a slow rise is often attributed to the acceleration of particles to higher energies, while a rapid decay is associated with efficient energy loss \citep{2024MNRAS.527.5140S}. Conversely, flares with rise times shorter than decay times are typically interpreted as the result of rapid particle injection followed by gradual radiative cooling \citep{2010ApJ...722..520A}. The negative asymmetry observed in our study aligns with the former physical scenario.

The variability timescales derived above can be used to constrain the characteristic size and location of the $\gamma$-ray emitting region. Assuming a spherical geometry, the size of the $\gamma$-ray emitting region can be constrained by the relation derived by \citet{1999A&AS..135..477R}:
\begin{equation}
R \le \frac{c \, t_{\mathrm{ob}}^{\mathrm{min}} \, \delta}{1 + z},
\label{eq:7}
\end{equation}
where $c$ and $z$ denote the speed of light and the redshift, respectively. Assuming a standard jet geometry, the distance from the emission region to the central supermassive black hole can be estimated using the observed variability timescale:
\begin{equation}
R_\mathrm{H} \le \frac{2 c \, t_{\mathrm{ob}}^{\mathrm{min}} \, \Gamma^2}{1 + z},
\label{eq:8}
\end{equation}
where $\Gamma$ represents the bulk Lorentz factor. Adopting bulk Lorentz factors of $\Gamma = 14.9$ and $\Gamma = 14.8$ (see Section \ref{sec_sed}) and assuming $\delta \approx \Gamma$ \citep{2011ApJ...733L..26A}, the sizes of the emission region during the F1 and F2 intervals are constrained to $R \le 2.2 \times 10^{15} \, \mathrm{cm}$ and $R \le 2.0 \times 10^{15} \, \mathrm{cm}$, with the corresponding distances from the central black hole limited to $R_\mathrm{H} \le 6.6 \times 10^{16} \, \mathrm{cm}$ and $R_\mathrm{H} \le 5.9 \times 10^{16} \, \mathrm{cm}$, respectively.

Furthermore, we found tentative evidence for flux variability in the minute-binned $\gamma$-ray light curve of BL Lacertae: during the major $\gamma$-ray flare on MJD 60588 (5 October 2024), the shortest flux doubling time was estimated to be $\tau = 0.7 \pm 0.2 \, \mathrm{min}$. If the shortest minute-scale variability timescale is adopted, the corresponding upper limit on the characteristic size of the variable emitting region would be $R \le 1.8 \times 10^{13}\,\mathrm{cm}$. This result is stringent than the $2.69 \times 10^{13}\,\mathrm{cm}$ limit derived by \citet{2022A&A...668A.152P} for the bright $\gamma$-ray flare of BL Lacertae on MJD 59331 (27 April 2021). Such an extremely compact scale is more naturally associated with localized substructures (e.g., plasmoids or mini-jets) within the relativistic jet, rather than with the entire $\gamma$-ray emitting region. The magnetic reconnection model featuring ultra-relativistic outflows can account for the extreme compactness of such small-scale emission regions and the observed sub-minute ultra-fast variability \citep[e.g.,][]{2009MNRAS.395L..29G,2024MNRAS.533..120C}. Within the magnetohydrodynamic (MHD) framework, a fraction of the jet's magnetic energy is dissipated through reconnection events, during which plasmoids are ejected from the reconnection zone at relativistic speeds, powering high-energy flares via inverse Compton scattering \citep{2009MNRAS.395L..29G,2020MNRAS.494.1817D}. This process can also account for longer variability timescales, on the order of hours, where a major flare may result from the superposition of emission from multiple independent plasmoids. At larger distances along the jet, the expansion of the emission region and the increase in electron cooling timescales can both contribute to longer variability timescales. The shortest variability may originate from reconnection zones associated with magnetic polarity reversals within compact regions of the jet. In contrast, longer-term variability may be linked to magnetic dissipation processes triggered by current-driven instabilities further downstream \citep{2009MNRAS.395L..29G}.

\subsection{The $\gamma$-ray Spectral Variability}
\label{sec_Spec}
The correlation between flux and photon index ($\Gamma_{\mathrm{ph}}$) serves as a powerful tool for tracing spectral evolution and probing the physical mechanisms underlying blazar variability. We therefore investigated the relationship between the $\gamma$-ray flux and the photon index. As shown in Figure \ref{f_4}, the 2-, 3- and 5-minute binned light curves obtained around MJD 60588.6 (5 October 2024) exhibit a ``softer-when-brighter'' trend, with Spearman's rank correlation coefficients of $r = 0.74$ ($p = 9.5 \times 10^{-4}$), $r = 0.66$ ($p = 0.02$), and $r = 0.96$ ($p = 4.5 \times 10^{-4}$), respectively. This finding is consistent with the minute-timescale spectral evolution (MJD 59331.2; 27 April 2021) reported by \citet{2022A&A...668A.152P}. More complex spectral behaviors have also been reported in BL Lacertae. \citet{2021MNRAS.507.5602P} reported a ``softer-when-brighter'' trend during an extreme $\gamma$-ray outburst over MJD 59060-59260 (30 July 2020 -- 15 February 2021), whereas a subsequent secondary flare exhibited a ``harder-when-brighter'' behavior. In addition, \citet{2024ApJ...967...96Y} performed a systematic analysis of short-timescale $\gamma$-ray flares from 29 high-Galactic-latitude BL Lac objects. They found that the flux-photon index distribution exhibits either ``harder-when-brighter'' or ``softer-when-brighter'' trends, but tends to flatten ($\Gamma_{\mathrm{ph}} \approx 2$) in the high-flux regime. Such differences in spectral evolution suggest that flares with distinct evolutionary patterns may originate from substantially different physical environments \citep{2021MNRAS.507.5602P} and/or be driven by the interplay between electron acceleration and radiative cooling processes \citep{2024ApJ...967...96Y}. Notably, the timing of the observed spectral softening coincides with the flare peak (MJD 60588.6; 5 October 2024) derived from a double-exponential fit, suggesting a close connection between the spectral evolution and the primary energy dissipation processes within the flare region.

Furthermore, Figure \ref{f_4} shows a saturation effect at high fluxes. \citet{2012ApJ...753...45S} also found that the photon indices of BL Lac objects become softer with increasing flux and eventually approach a nearly constant value above a certain flux threshold. \citet{2024ApJ...967...96Y} found that the flux--photon index relation exhibits a truncated feature toward the high-flux state, where the photon index becomes nearly constant. The observed saturation of the photon index at high fluxes may indicate that the electron energy distribution approaches a steady state, in which the competition between particle acceleration and radiative cooling inside the jet reaches a dynamic equilibrium, thereby limiting further spectral evolution \citep{2024ApJ...967...96Y}.

\subsection{Broadband SED Modeling}
\label{sec_sed}
We employ a one-zone leptonic model to fit the broadband SED. As illustrated in Figure \ref{f_5}, the low-energy peak of the SED is dominated by synchrotron radiation, whereas the X-ray emission originates predominantly from the synchrotron self-Compton (SSC) process. The high-energy peak is reproduced by the external Compton (EC) process. The model results are displayed in Figure \ref{f_5}, and the corresponding parameters are summarized in Table \ref{table3}. It should be noted that, in the SEDs of the F2 and low state, the two highest-energy $\gamma$-ray data points are not well reproduced by the best-fit model. However, these points have relatively low detection significances, with TS values of approximately 20, and their predicted counts ($N_{\rm pred}$) are both below 7. By contrast, all other $\gamma$-ray data points have TS values exceeding 100. Owing to their comparatively large uncertainties, the two highest-energy points carry relatively low statistical weight in the fitting procedure and therefore do not significantly affect the overall goodness of fit. The resulting reduced $\chi^2$ values are 1.1, 0.9, and 1.6 for the F1, F2, and low states, respectively, indicating that the adopted model still provides an acceptable description of the broadband SEDs despite the discrepancies at the highest energies.

The SED modeling results indicate that the emission region sizes for flares F1 and F2 are $1.5 \times 10^{15}$ cm and $1.7 \times 10^{15}$ cm, respectively, with corresponding locations at $2.4 \times 10^{16}$ cm and $3.3 \times 10^{16}$ cm. These values are consistent with the upper limits inferred from the observed variability timescales. In this study, the derived jet bulk Lorentz factors during the flaring states are 14.9 and 14.8, which are smaller than the values of 18.45 - 20.09 reported by \citet{2024MNRAS.527.5140S} for the 2021 - 2022 flaring episode. The injected electron spectral index in the low state is $p_{1}=2.5$, close to the value of $p_{1}=2.38$ reported by \citet{2024MNRAS.527.5140S} for the quiet phase. In contrast, the injected electron spectral indices during the F1 and F2 states are $p_{1}=1.2$ and $1.5$, respectively, which are comparable to the value of $p_{1}=1.65$ obtained by \citet{2025JHEAp..4800402M}. The harder $p_{1}$ during the flaring states suggests a more efficient particle acceleration process. Such hard spectra can be naturally explained by relativistic magnetic reconnection, during which magnetic energy within the jet is efficiently dissipated and converted into particle kinetic energy \citep{2009MNRAS.395L..29G}. In particular, this mechanism is capable of producing exceptionally hard particle injection spectra with indices as low as $p\lesssim1.5$ \citep{2014ApJ...783L..21S}.

Our SED modeling for the low state yields a magnetic field strength of 2.3~G. During the flaring states, the magnetic field strengths are 0.5 and 0.4~G, respectively. These values are consistent with the range of 0.50 - 0.75~G reported by \citet{2024ApJ...974..233K}. Similarly, \citet{2024MNRAS.527.5140S} found that the magnetic field strength of BL Lacertae ranged from 0.49 to 0.72~G during different flaring episodes. Assuming that the flares are driven by particle acceleration via magnetic reconnection, during which magnetic energy is efficiently transferred to particles and plasmoids are formed, \citet{2023MNRAS.521L..53A} derived a magnetic field strength of $\sim 0.6\, \mathrm{G}$ within a reconnection region located near the outer boundary of the broad-line region ($\sim0.02\,\mathrm{pc}$), which is also in agreement with our results. Taken together, the hard injected electron spectra and the relatively low magnetic field strengths inferred during the flaring states are consistent with a scenario in which magnetic reconnection plays an important role in powering the observed broadband SED emission.

Furthermore, since the EC emission depends on the external photon field provided by the BLR, variations in $L_{\text{disk}}$ can affect the SED modeling. Such variations influence the intrinsic luminosity of the BLR and determine its characteristic radius through the relation $R_{\text{BLR}} \propto L_{\text{disk}}^{1/2}$, thereby affecting the contribution of the EC process to the high-energy component of the SED \citep{2011MNRAS.414.2674G}. To assess this effect, we perform additional SED fits by adopting $L_{\rm disk} = 1.4 \times 10^{43}$ and $3.4 \times 10^{43}\ {\rm erg\,s^{-1}}$, while keeping other model assumptions unchanged. The resulting reduced $\chi^2$ values remain statistically acceptable, with values of approximately 1.4, 1.1, and 1.0 for the low, F1, and F2 states, respectively. The derived jet parameters show only moderate variations, indicating that our main SED modeling results remain robust within the tested $L_{\rm disk}$ range. At sufficiently low $L_{\rm disk}/L_{\rm Edd}$, the accretion flow may enter a radiatively inefficient regime, potentially reducing its ability to photo-ionize the broad-line clouds \citep{2009MNRAS.397..985G}. In addition, as the viewing angle ($\theta$) is a critical parameter affecting Doppler boosting, we further investigate its influence on the broadband SED modeling. Because the viewing angle is not a directly measured quantity, it may introduce uncertainties into the derived jet parameters. Therefore, while keeping all other model assumptions unchanged, we perform supplementary fits by fixing $\theta$ at $0.1^\circ$, $0.2^\circ$, $0.5^\circ$, and $1.0^\circ$. Across all tested cases, the resulting reduced $\chi^2$ values remain statistically acceptable, ranging from $1.4\text{--}1.6$ for the low state, $1.1\text{--}1.2$ for the F1 state, and $\sim 1.0$ for the F2 state. These results indicate that broadband SED data alone cannot tightly constrain the viewing angle within the sub-degree regime. Although variations in $\theta$ lead to changes in certain derived jet parameters, the overall SED shape and physical interpretations remain robust. Consequently, the adopted value of $\theta = 0.1^\circ$ should be regarded as a representative value consistent with previous BL Lacertae studies, rather than a uniquely constrained parameter.

\section{Summary}
\label{sec_sum}
\vspace{0.4in}
In this work, we present a detailed multi-wavelength analysis of the blazar BL Lacertae during the exceptionally active states between MJD 60500 and 60800 (9 July 2024 -- 5 May 2025). By employing Bayesian Blocks and HOP algorithms on \textit{Fermi}-LAT data, we identified two primary $\gamma$-ray flaring episodes (F1 and F2). Our main results are summarized as follows:

(1) Using the \textit{Fermi}-LAT orbit-binned light curves, we derived a minimum flux halving timescale of $1.33 \pm 0.29$ hr, constraining the size of the $\gamma$-ray emitting region to $R \le 2.0 \times 10^{15}$ cm. We also found tentative evidence for sub-minute $\gamma$-ray variability in the 2-minute binned light curves, with a shortest flux doubling timescale of $\tau = 0.7 \pm 0.2$ min. This rapid variability implies an extremely compact emission region ($R \le 1.8 \times 10^{13}$ cm), suggesting that the observed emission may originate from localized magnetohydrodynamic structures, such as plasmoids within a magnetic reconnection zone.

(2) A significant ``softer-when-brighter'' trend is identified during the flare peaks at minute scales. The correlation between the \textit{Fermi}-LAT flux and the power-law photon index ($r = 0.96$, $p = 4.5 \times 10^{-4}$) suggests that the spectral evolution is intimately linked to the primary energy dissipation processes and the interplay between particle acceleration and radiative cooling.

(3) The broadband SEDs are successfully reproduced within a one-zone leptonic framework. The low-energy component is dominated by synchrotron emission, while the X-ray emission is primarily attributed to the SSC process, and the high-energy peak is well explained by EC scattering. The reduced magnetic field strengths and hard electron injection spectral indices during the flaring states suggest enhanced particle acceleration efficiency, possibly associated with relativistic magnetic reconnection.

\section*{Acknowledgements}
This work is supported by the National Key Research and Development Program of China (No. 2024YFA1611603) and the Yunnan Key Laboratory of Survey Science (No. 202449CE340002). Mephisto is developed at and operated by the South-Western Institute for Astronomy Research of Yunnan University (SWIFAR-YNU), funded by the ``Yunnan University Development Plan for World-Class University'' and ``Yunnan University Development Plan for World-Class Astronomy Discipline''. This work has made use of \textit{Fermi} data from the \textit{Fermi} Science Support Center (FSSC) and \textit{Swift} data from the High Energy Astrophysics Science Archive Research Center (HEASARC), both supported by NASA's Goddard Space Flight Center (GSFC). This work is also supported by the National Natural Science Foundation of China (grants 11863007, 12063005, 12063007, 11703078, 12473020), the Yunnan Province Foundation (2019FB004), the Yunnan Fundamental Research Projects (grant No. 202601AT070222), the Program for Innovative Research Team (in Science and Technology) in University of Yunnan Province (IRTSTYN), and Yunnan Local Colleges Applied Basic Research Projects (2019FH001-12). We acknowledge the science research grants from the China Manned Space Project with No. CMS-CSST-2021-A06. The authors acknowledge support from the ``Science \& Technology Champion Project'' (202005AB160002) and from two ``Team Projects'' - the ``Innovation Team'' (202105AE160021) and the ``Top Team'' (202305AT350002), all funded by the ``Yunnan Revitalization Talent Support Program''.

\section*{Data availability}
The data underlying this article are provided the online journal.

\end{document}